\documentclass[twocolumn,prb]{revtex4}
\usepackage{amsmath}
\usepackage{graphicx}
\usepackage{epsfig}
\newlength{\upit}\upit=0.1truein
\newcommand{\zmatrix}[4]{\left(\begin{matrix}#1 & #2\cr #3&#4\end{matrix}\right)}

\newcommand{\ltappr}{{{\lower4pthbox{$<$} } \atop \widetilde{ \ \ \ }}}
\newlength{\bxwidth}\bxwidth=1.5 truein

\newcommand{\tr}{{\hbox{Tr}}}

\newcommand{\dg}{^{\dagger }}

\newcommand{\bS}{\mathbf{S}}
\newcommand{\rarrow}{\rightarrow}
\newcommand \bea {\begin{eqnarray} }
\newcommand \eea {\end{eqnarray}}

\newcommand{\bx}{{\bf{x}}}
\newcommand{\calS}{\mathcal{S}}
\newcommand{\calT}{\mathcal{T}}
\newcommand{\calP}{\mathcal{P}}

\newlength{\figwidth}
\newlength{\shift}
\newcommand{\fg}[3]
{
\begin{figure}[ht]

\vspace*{-0cm}
\[
\includegraphics[width=\figwidth]{#1}
\]
\vskip -0.2cm
\caption{\label{#2}
\small#3
}
\end{figure}}

\begin{document}

\title{Symplectic N and time reversal in frustrated magnetism
}

\author{Rebecca Flint and P. Coleman}
\affiliation{
Center for Materials Theory,
Rutgers University, Piscataway, NJ 08855, U.S.A. 
} 

\begin{abstract}
Identifying the time reversal symmetry of spins as a symplectic
symmetry, we develop a large $N$ approximation for quantum magnetism
that embraces both antiferromagnetism and ferromagnetism.  In $SU(N)$,  $N>2$, not all spins invert under time reversal,
so we have introduced a new large $N$ treatment which builds interactions exclusively out of the symplectic subgroup[$SP(N)$] of time reversing spins, a more stringent condition than
the symplectic symmetry of previous $SP(N)$ large $N$ treatments.  As a result, we obtain a mean field theory that incorporates the energy cost of frustrated bonds.  When applied to the frustrated square lattice, the ferromagnetic bonds 
restore the frustration dependence of the critical spin in the N\'eel phase, and recover the correct frustration dependence of the finite temperature Ising transition.
\end{abstract}

\maketitle

\section{Introduction}
The search for simple, controlled approximations which capture the collective behavior of matter is a key goal of condensed matter.  In quantum magnetism, this search is hindered by the lack of a small parameter; after more than a decade, theorists and experimentalists are still searching for a physically realizable quantum spin liquid\cite{sliquid}, and the ground state behavior of highly frustrated magnets, like the kagom\'e\cite{pyrochlore, sachdevkagome,kagomeexpt}, pyrochlore\cite{oleg,pyrochlore} and hyperkagom\'e\cite{takagi,hyperkagomeclassical,ybkim} lattices is still unclear.  One approximation that has proven successful is the ``large $N$'' expansion, which generalizes the model of interest to a family of models where the number of internal degrees of freedom is indexed by an integer $N$.  As $N$ goes to infinity, central limit effects permit the underlying collective behavior of the model to be solved exactly, and finite $N$ properties may be obtained from a power series expansion in $1/N$ about this solution.  

The basic equation of quantum magnetism is the Heisenberg Hamiltonian,
\begin{equation}
H = \sum_{ij} J_{ij} \vec S_i \cdot \vec S_j
\end{equation}
where the spin on each site,  $\vec S_i$ lives in the group $SU(2)$.  The exchange coupling $J$ can be either positive or negative, for simple lattices these lead to antiferromagnetic or ferromagnetic ground states, respectively.  Both ground states break both spin rotational and time reversal symmetries, but the antiferromagnet is invariant under the combination of time reversal and  translation by one lattice site.  More complicated lattices can lead to spins which are not collinera, so called spiral magnets, or possibly to a state in which the spins are not ordered at all, a spin liquid. 

As we extend the theory of interacting $SU(2)$ spins into a family of related theories, we will lose some of the physics unique to $SU(2)$ spins, 
so how do we guarantee that our resulting theories still capture the defining characteristics of magnetism?  What defines magnetism?  What 
defines a spin?  As always, the first, best place to look is at the symmetries.  An $SU(2)$ spin Hamiltonian has two symmetries - time reversal invariance and invariance with respect to $SU(2)$ rotations.  XY and Ising spin Hamiltonians also obey time reversal and rotational invariance, but  under $U(1)$ or $Z_2$ rotations.  The spins themselves define a unique direction on a manifold,  $CP^1$ for $SU(2)$, and invert under time reversal, 
$\vec S \rarrow -\vec S$.  
The ground state can break the rotational and time reversal symmetries, traditionally simultaneously, as in a ferromagnet, but more recently hypothesized states can break either rotational symmetry, but not time inversion; e.g. a spin-nematic defines a unique direction, but does not have magnetic long range order\cite{qsnematic}, or chiral spin states which break time reversal, but not rotational symmetry\cite{chiralspin}.
Certainly there are nonmagnetic states, such as liquid crystal displays, which also break rotational symmetry, so the rotational properties of spin are not enough to define magnetism.  We propose that both the rotational and time reversal properties of spins are defining symmetries of magnetism, and that a large $N$ theory with broad applicability must maintain both of these properties in the large $N$ limit.

Large $N$ theories are well known in particle physics\cite{witten} and in heavy fermion theory\cite{read,read2,auerbach}, but they were first introduced in quantum magnetism by Berlin and Kac, who solved the spherical model of ferromagnetism exactly in a large $N$ limit\cite{spherical}.
Simultaneously, Anderson, Dyson and Maleev introduced spin wave theory, which takes the spin $S$ to be large and expands in $1/S$\cite{anderson,dyson,maleev}.  The large $S$ limit is a classical limit, where the spins behave like classical vectors, rotating under the group $O(3)$.  
The long wavelength fluctuations of quantum spins were studied semiclassically in the nonlinear sigma model\cite{NLSM}, where quantum renormalizations to the classical parameters were calculated in the large $N$ limit by extending the order parameter manifold of $SU(2)$ spins, $CP^1$ to that of $SU(N)$ spins, $CP^{N-1}$\cite{dadda}.

The large $N$ quantum limit of magnetism was first treated by Affleck and Marston\cite{affleckchains, affleckmarston}.  They used a fermionic spin representation to treat the $S=1/2$ Hubbard and Heisenberg models by extending $SU(2)$ to $SU(N)$, preserving the rotational invariance of the Hamiltonian under $SU(N)$.  
Since then, other extensions of $SU(2)$ have been used, including $SP(2N)$ by Ran and Wen\cite{wen}.
Fermionic large $N$ theories capture the physics in the extreme quantum limit, $S/N \ll 1$, where the ground state is always disordered. 
These are useful for studying the possible spin liquid ground states\cite{U1spinliquid}, but not for determining if a particular model is a spin liquid in the first place.  For that, one needs a bosonic spin representation, where magnetic long range order corresponds to the condensation of the bosons.  Arovas and Auerbach introduced the bosonic $SU(N)$ theory\cite{schwinger}, which can treat arbitrary ratios of $S/N$, and both magnetically ordered and disordered states.
This theory was quite successful at describing ferromagnets and bipartite antiferromagnets, but is unable to treat frustrated magnets.  To resolve this, Sachdev and Read extended the theory to arbitrary antiferromagnetic bonds by limiting the rotational invariance to the group $SP(N)$\cite{readsachdev91}.  However, neither of these theories preserve the time inversion properties of spins, because for $N>2$, not all $SU(N)$ spins invert under time reversal, and although Sachdev and Read's Hamiltonian is invariant under $SP(N)$ rotations, it still contains $SU(N)$ spins with  the wrong parity under time reversal.  
Recently we introduced a new large $N$ limit, which identifies time reversing spins with the generators of $SP(N)$, and then builds interactions exclusively from these symplectic spins\cite{nphysus}.  This condition is more stringent than Sachdev and Read's, and leads to a unique large $N$ limit which we call ``symplectic-$N$''.   In collaboration with Dzero, we introduced symplectic-$N$ using a fermionic spin representation to treat Kondo physics and superconductivity in the two channel Kondo model\cite{nphysus}.  Here, we develop the bosonic symplectic-$N$ approach for the Heisenberg model, which enables us to treat ferromagnetism and antiferromagnetism on equal footing.

The structure of this paper is as follows.  In section \ref{TimeRev}, we show that the time reversal of spins is a symplectic property, and extend time reversal to large $N$ where the $SU(N)$ generators separate into two classes - those that reverse under time reversal, and those that do nothing.   We examine different decouplings of the large $N$ Heisenberg hamiltonian and show that excluding the non-time reversing spins from the interaction Hamiltonian captures both ferromagnetic and antiferromagnetic correlations.  In section \ref{GenSympH}, we derive the mean field equations for a generic Heisenberg magnet in the symplectic-$N$ limit, while in section \ref{sectionJ1J2}, we apply these ideas to the two dimensional $J_1-J_2$ model, finding both the zero temperature and finite temperature phase diagrams.  Finally, in section \ref{Conclusions}, we draw conclusions about the application of symplectic-$N$ to other models.

\section{Time reversal and symplectic symmetry}\label{TimeRev}

Time reversal is defined by its action on an electron wavefunction $\psi_\sigma(\bx,t)$:
\begin{equation}
\theta \psi_\sigma(\bx,t)  = {\tilde \sigma} \psi_{-\sigma}^*(\bx,-t).
\end{equation}
More generally it is a matrix operator, $\theta = \hat \epsilon K$, where $K$ is the complex conjugation operator, $K \psi = \psi^* K$ and $\hat \epsilon$ is the antisymmetric matrix $i \sigma_2$.  A consistent definition of time reversal requires that $\theta$ commute with the unitary rotation operators $U$, the members of the group $SU(2)$, 
\begin{equation}
U \theta U\dg = \theta.
\end{equation}
Using the definition of $\theta = \hat \epsilon K$, and noting that $K$ converts $U\dg$ to $U^T$,
we find 
\begin{equation}
U \hat \epsilon U^T  =  \hat \epsilon
\end{equation}
This is a symplectic condition on the matrices $U$ because it requires the invariance of an antisymmetric matrix $\hat \epsilon$ under orthogonal transformations.
 If  $U$ is taken to be the matrix for an infinitesimal rotation, 
$U = 1+ {\vec \alpha} \cdot {\vec S}$, the symplectic condition requires that 
\begin{equation}
{\vec S} \rarrow \theta {\vec S} \theta^{-1} = \hat \epsilon {\vec S}^T \hat \epsilon = -{\vec S}.
\end{equation}
So the symplectic condition is equivalent to the time reversal of all $SU(2)$ spins.

\fg{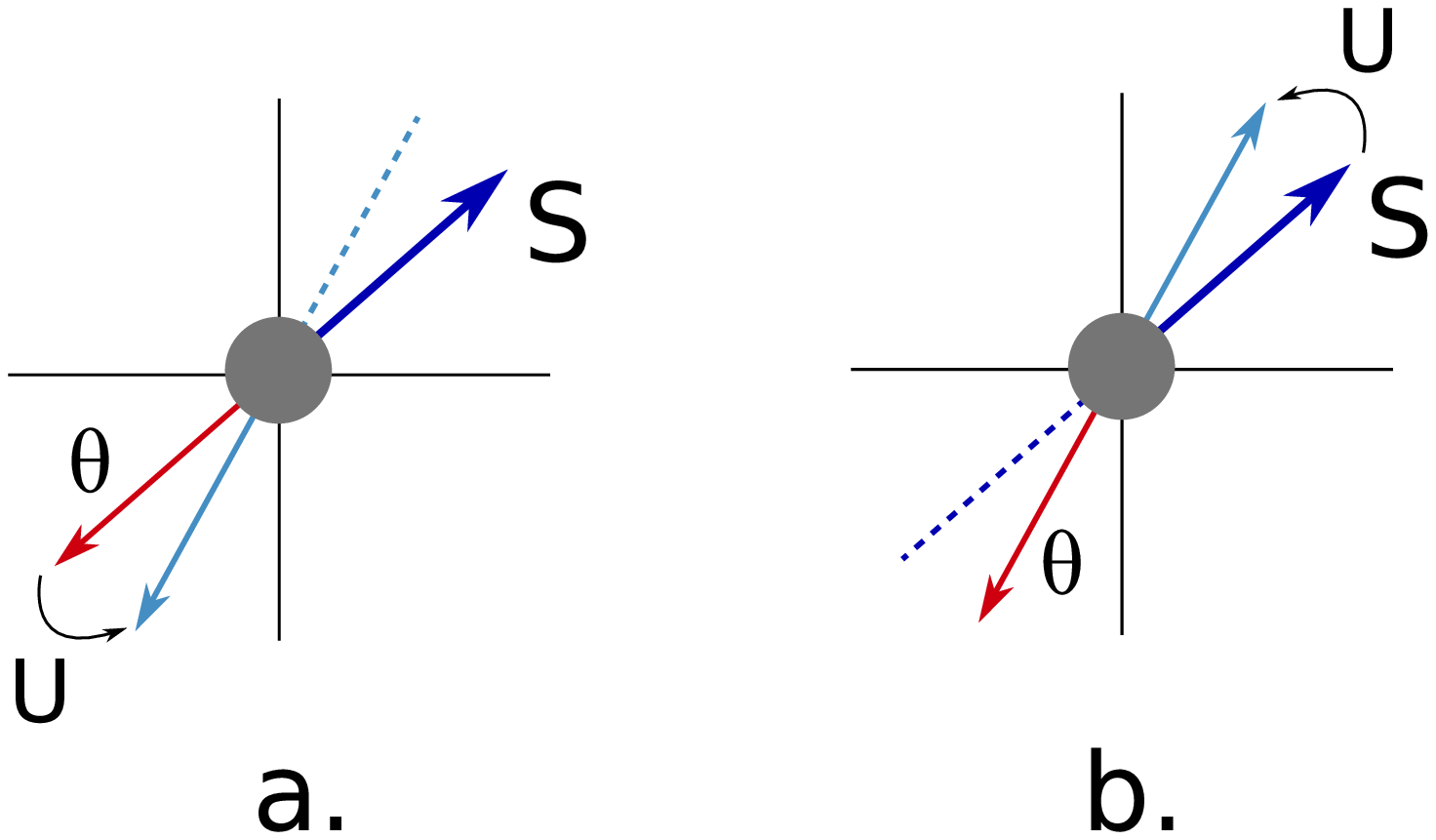}{commutefig}{The time reversal operator, $\theta$ and rotation operators $U$ acting on a spin must commute.  a. Depicts $\vec S \xrightarrow{\theta U} -R \vec S$.  b. $\vec S \xrightarrow{U\theta} R(- \vec S)$, where $R$ is the rotation performed by $U$.  These two are equivalent for all $U$ in $SU(2)$, and $SP(N)$, but not in $SU(N)$.}

\subsection{Time reversal in large $N$}

In quantum magnetism, it is convenient to use the Schwinger boson representation of spins\cite{schwinger},
\begin{equation}
\hat S^a_j = \frac{1}{2}b\dg_{j, \alpha} S^a_{\alpha\beta} b_{j, \beta}
\end{equation}
where $b\dg_{j} = (b\dg_{j,+1}, b\dg_{j,-1})$ is a two component spinor defined on each site and $S^a$ is one of the $SU(2)$ generators, eg - the Pauli matrices.  When this treatment is generalized to large $N$, the number of spin components increases from 2 to an even number $N = 2k$. Dropping the site index, we have
$\hat \calT^a_j = \frac{1}{2}b\dg_{j, \alpha}\calT^a_{\alpha\beta} b_{j, \beta}$, 
where 
\begin{equation}
b\dg = \left(b\dg_{+1}, b\dg_{-1}, b\dg_{+2}, b\dg_{-2}, \cdots, b\dg_{+k}, b\dg_{-k}\right)
\end{equation}
and $\calT^a$ are the generators of the group $SU(N)$.

Time reversal is a defining property of magnetism, so when we extend the number of spin components, we want to maintain this
essential discrete symmetry.  However, $SU(N)$ generators divide into two classes under time reversal(Fig \ref{SympPlane}),
\begin{equation}
\hat \epsilon \left(\calT^a\right)^T \hat \epsilon = \left\{ \begin{array}{lcl}
- \calT^a && a \in \{1,2,\dots, D_{N} \} \cr
+ \calT^a && a \in \{D_{N}+1,\dots, N^2+1\}\cr
\end{array}\right. , 
\end{equation}
where $D_{N} = \frac{1}{2}N (N+1)$.  The first class can be identified as the generators of the symplectic subgroup, $SP(N)$, whose elements reverse under time reversal, just like the $SU(2)$ spins.  $D_{N}$ is the number of $N$ dimensional symplectic generators. To avoid confusion, we will label these symplectic spins by $\mathcal{S}^a$.   The second class does not invert under time reversal, and does not form a closed subalgebra of $SU(N)$.  By analogy with the fermionic representation\cite{nphysus}, we know that these ``antisymplectic'' generators change sign under particle-hole transformations, or charge conjugation, while the symplectic spins are neutral.  These generators then behave like \emph{electric} dipoles, not like magnetic moments, so we label them by $\mathcal{P}^a$.  For $N=2$, $SU(2) \cong SP(2)$, and there are no antisymplectic generators.  However, for any $N>2$, the two groups are no longer isomorphic.
For example, $SU (4)$ consists of ten symplectic generators 
\begin{equation}\label{}
\calS^a \in \left\{
\zmatrix{}{i\underline{1}}{-i\underline{1}}{},
\zmatrix{}{\vec{\sigma }}{\vec{\sigma}}{},
\zmatrix{\vec{\sigma }}{}{}{\pm\vec{\sigma }}
\right\}
\end{equation}
(corresponding to the four Dirac matrices $\gamma_{\mu}$ and their six
commutators $\frac{i}{2}[\gamma_{\mu},\gamma_{\nu }]$),
and five antisymplectic generators
\begin{equation}\label{}
\calP^a \in \left\{
\zmatrix{\underline{1}}{}{}{-\underline{1}},
\zmatrix{}{i\vec{\sigma }}{-i\vec{\sigma}}{},
\zmatrix{}{\underline{1}}{\underline{1}}{}
\right\}.
\end{equation}
(corresponding to the $\gamma_{5}$ matrix, and its product with the
four Dirac matrices $i \gamma_{5 }\gamma_{u}$.)

Here, the choice of $SP(N)$ is motivated by the desire to maintain the time reversal symmetry of spin in the large $N$ limit, but Sachdev and Read originally developed $SP(N)$ because, unlike $SU(N)$ it contains well defined particle-particle singlets\cite{readsachdev91}.  $SU(N)$ expansions are extremely useful to particle physicists because there are two well defined color singlets - mesons and baryons.  Mesons are particle-antiparticle pairs, or in condensed matter, particle-hole pairs, while baryons are products of $N$ particles, forming the three quark baryons for $SU(3)$, which have no condensed matter analog except for $N=2$, where these are particle-particle pairs - e.g.  valence bonds\cite{valencebonds} or Cooper pairs.  In the large $N$ limit, the condensed matter version of $SU(N)$ has only particle-hole pairs.  However, the group $SP(N)$ does have well defined particle-particle singlets, which are the pairing of a particle and its time reversed twin, and particle-hole pairs, but no baryons  The presence of these well defined singlets is equivalent to the existence of a well defined time reversal symmetry of spin.

\subsection{Decouplings of the Heisenberg Hamiltonian}

In order to treat magnetic interactions, we would like to rewrite the Heisenberg Hamiltonian, $H = J \hat \calS_i \cdot \hat \calS_j$ without explicit reference to the spin generators. In $SU(N)$, this is done by using the $SU(N)$ completeness relation: 
\begin{equation}
\label{SUNcomplete}
\sum_a \mathcal{T}^a_{\alpha \beta} \cdot \mathcal{T}^a_{\gamma \eta} = 2 \delta_{\alpha \eta} \delta_{\beta \gamma} - \frac{2}{N} \delta_{\alpha \beta} \delta_{\gamma \eta}
\end{equation}

We now derive a similar $SP(N)$ completeness relation.  Any even dimensional matrix can be split into a symplectic and antisymplectic part: $M = M_S + M_A$, where the symplectic part satisfies $M_S = -\hat \epsilon M_S^T \hat \epsilon^T$ and the antisymplectic part $M_A = \hat \epsilon M_A^T \hat \epsilon^T$.  The symplectic part can be obtained by projection, $M_S ={\rm P} M$, where ${\rm P}$ is defined such that ${\rm P} M_A = 0$.  We recognize that $M_A - \hat \epsilon M_A^T \hat \epsilon^T = 0$, and take
\begin{equation}
{\rm P} M = \frac{1}{2}\left(M - \hat \epsilon M^T \hat \epsilon^T\right)
\end{equation}
This expression can be written out in terms of components, 
\begin{eqnarray}\label{l}
{\rm P}^{\alpha \beta }_{\gamma\eta}M^{\eta \gamma} &=& \frac{1}{2} 
[M^{\alpha \beta }- {\epsilon }^{\alpha }_{\gamma}M^{\eta \gamma} \epsilon^{\beta }_{\eta  }]\cr
&=& 
\frac{1}{2} 
[\delta^{\alpha }_{\eta }\delta^{\beta }_{\gamma}- { \epsilon }^{\alpha }_{\gamma}
 \epsilon^{\beta }_{\eta }]
M^{\eta \gamma},
\end{eqnarray}
so that
\begin{equation}\label{projection}
{\rm P}^{\alpha \beta}_{\gamma\eta } = 
\frac{1}{2} 
[\delta^{\alpha }_{\eta}\delta^{\beta }_{\gamma}- { \epsilon }^{\alpha }_{\gamma}
 \epsilon^{\beta }_{\eta }].
\end{equation}
Since the symplectic matrices form a group,  $M_S$ can always be expanded in the symplectic generators, $\calS^a$, $M_S = \sum_a m_a \mathcal{S}^a$.  With the normalization $\tr \left[\mathcal{S}^a \mathcal{S}^b \right]=2 \delta_{a b}$, consistent with the $SU(2)$ Pauli matrices, the coefficient $m_a = \frac{1}{2} \tr \left[\mathcal{S}^a M\right]$, giving ${\rm P} M = \frac{1}{2} \sum_a \tr \left[ \mathcal{S}^a M\right] \mathcal{S}^a$.
Expanding both sides in terms of components and canceling $M^{\eta \gamma}$, we find
\begin{equation}\label{}
{\rm P}^{\alpha\beta }_{\gamma\eta }=
\frac{1}{2}\sum_{a}\mathcal{S} ^{a}_{\alpha \beta }\mathcal{S}^{a}_{\gamma\eta }.
\end{equation}
Finally, by inserting (\ref{projection}), we obtain the $SP(N)$ completeness relation,
\begin{eqnarray}\label{complete}
\sum_{a} (\mathcal{S}^{a})_{\alpha \beta } (\mathcal{S}^{a})_{\gamma \eta }=
[\delta^{\alpha }_{\eta }\delta^{\beta }_{\gamma}- { \epsilon }^{\alpha }_{\gamma}
 \epsilon^{\beta }_{\eta }].
\end{eqnarray}

Inserting the  Schwinger boson spin representation,
the symplectic $N$ Heisenberg Hamiltonian becomes the sum of two terms
\begin{equation}\label{spinham}
\hat \calS_{i}\cdot \hat \calS_{j} =
- B\dg_{ji}B_{ji}+ A\dg_{ji}A_{ji}
\end{equation}
where  
\begin{equation}
B\dg _{ji} 
=\frac{1}{2}\sum_{\sigma }\tilde{\sigma } b \dg_{j\sigma }b \dg _{i-\sigma }
\end{equation}
creates a valence bond, or spin singlet pair, between sites i and j, and
\begin{equation}
A\dg_{ji}=
\frac{1}{2}\sum_{\sigma }b \dg_{j\sigma }b _{i\sigma }
\end{equation}
creates a ferromagnetic bond, which imply the coherent hopping of Schwinger bosons from site to site.  In the language of valence bonds, a ferromagnetic bond can be 
thought of as resonating one end of a valence bond between sites $i$ and $j$, causing both sites
to be simulataneously antiferromagnetically correlated with a third site, thus ferromagnetically correlated with one another.  In this sense, it is a frustrating field.  Most generally, a ferromagnetic bond on a link with antiferromagnetic $J$, or vice versa, can be considered frustrating fields, however, we will usually be dealing with entirely antiferromagnetic lattices, where any ferromagnetic bond is a frustrated bond.
This decoupling is identical to the $SU(2)$ mean field theory introduced by Ceccatto et al\cite{ceccatto}, now controlled by the large $N$ limit of properly time reversing spins.
Next, we compare this representation with $SU(N)$\cite{schwinger} and the previous $SP(N)$ treatment\cite{readsachdev91}.

The $SU(N)$ Heisenberg Hamiltonian is a dot product between $SU(N)$ spins, $\hat \calT$, which can
be rewritten using the $SU(N)$ completeness relation(\ref{SUNcomplete}) to obtain the usual sum of ferromagnetic bonds\cite{schwinger}
\bea
H_{SU(N)} & = & \frac{J_{ij}}{N} \hat \calT_i \cdot \hat \calT_j = \frac{2 J_{ij}}{N} A\dg_{ji} A_{ji}\\
 & = &  \frac{J_{ij}}{N} \left(\hat \calP_i \cdot \hat \calP_j  +\hat \calS_i \cdot \hat \calS_j \right),
\eea
where $J_{ij}$ is rescaled by $N$ so that $H$ is extensive in $N$.
As one would expect in $SU(N)$, the symplectic and antisymplectic spins are treated on equal footing, which leads to a completely ferromagnetic theory.  Bipartite antiferromagnets can also be studied in $SU(N)$ by performing a special transformation(not time reversal) on one sublattice, but $SU(N)$ cannot treat more complicated, e.g.- frustrated, antiferromagnets.

The $SP(N)$ Hamiltonian, as defined by Sachdev and Read\cite{readsachdev91} was originally written in terms of valence bonds, $H_{SP(N)} = -J_{ij} B\dg_{ji} B_{ji}$, in order to treat frustrated antiferromagnets.  When we rewrite it in terms of magnetic and electric dipoles, we find
\bea
H_{SP(N)} & = & - \frac{J_{ij}}{N} B\dg_{ji} B_{ji} \\
 & = & \frac{J_{ij}}{2N}\left(\hat \calS_i \cdot \hat \calS_j - \hat \calP_i \cdot \hat \calP_j\right).
\eea
Surprisingly, the $SP(N)$ large $N$ theory weights the physical symplectic and unphysical antisymplectic spins equally, but with opposite signs.  $SP(N)$ was so called because the Hamiltonian satisfies symplectic symmetry, not because it describes the interactions of symplectic spins.  In fact, any combination of the two terms $B\dg B$ and $A\dg A$ has symplectic symmetry, including $SU(N)$.  {\bf The requirement that our interactions include only magnetic, symplectic spins is more stringent, and this method is what we call symplectic-$N$}, while we will continue to refer to Sachdev and Read's formulation as $SP(N)$.

\vspace{3mm}

\begin{center}
\begin{tabular}{|c|c|c|}
\hline
Approach & H($\calS,\calP$) & H($b\dg,b$)\\
\hline
$SU(N)$ & $J\left( \calS \cdot \calS +\calP \cdot \calP\right)$ & $J A\dg A$\\
$SP(N)$ & $J\left( \calS \cdot \calS - \calP \cdot \calP\right)$ & $-J B\dg B$\\
{\bf Symplectic-$\mathbf{N}$} & $J \calS \cdot \calS$ & $ J\left( -B\dg B +A\dg A\right)$\\
\hline
\end{tabular}
\end{center}

\vspace{3mm}

Why is it important to exclude the non-time reversing dipoles?  Both the symplectic ($\hat \calS_i \cdot \hat \calS_j$) and antisymplectic ($\hat \calP_i \cdot \hat \calP_j$) interactions are invariant under time reversal, however, the important difference is not in the Hamiltonian, but in the ground states, and the dynamics.  These are far more coupled than the Hamiltonian suggests because the $SU(N)$ spin $\hat \calT$ does not act as a vector, and the antisymplectic and symplectic directions are not independent directions,  so that $\hat \calT$ is unable to point in a purely symplectic direction.  The antisymplectic interactions encourage the antisymplectic spins to order - competing with the ordering of the physical components.  This competition eliminates the antiferromagnetic[ferromagnetic] ground state completely for $SU(N)$[$SP(N)$].  And finally, even if the ground state is the one of interest, the presence of antisymplectic interactions affects the dynamics of the symplectic spins, dynamically violating the closure of the symplectic subgroup.

\subsection{Constraints and spin Casimirs}\label{casimir}

In the Schwinger boson representation, the total spin on site is free to take any value.  Therefore, in order to treat a real spin problem, the total spin must be restricted to its physical value, ${\vec S}^2 = S(S+1)$.  This contraint is implemented by a Lagrange multiplier fixing the value of the spin Casimir $(\vec S_j)^2$, which depends on the group.  For a general group with generators $\Gamma^a$, the Casimir is written,
\begin{equation}
\hat \bS_j ^{2}= \sum_a
(
\frac{1}{2}b\dg_{j\alpha } \Gamma ^{a}_{\alpha \beta }b_{j\beta })
(\frac{1}{2}b\dg_{j\gamma} \Gamma^{a}_{\gamma\eta }b_{j\eta }).
\end{equation}
For symplectic spins, the completeness relation(\ref{complete}) is used to rewrite the Casimir as
\bea
\hat \calS_j ^{2}&=& \frac{1}{4}\left(b\dg_{j\alpha} b_{j\beta}\right)\left(b\dg_{j\gamma} b_{j\eta}\right)\left[\delta_{\alpha\eta}\delta_{\beta \gamma} + \epsilon_{\alpha \gamma} \epsilon_{\eta \beta}\right] \cr
& = & \frac{1}{4} \left( b\dg_{j\alpha}b_{j\beta}b\dg_{j\beta}b_{j\alpha} + \tilde{\alpha}\tilde{\beta}b\dg_{j\alpha}b_{j-\beta} b\dg_{j-\alpha} b_{j\beta}\right)\cr
& = & \frac{1}{4}\left(\left[ b\dg_{j\alpha}b_{j\alpha} b_{j\beta}b\dg_{j\beta}-n_{bj}\right]
+\left[\tilde{\alpha}\tilde{\beta}b\dg_{j\alpha}b\dg_{j-\alpha}b_{j-\beta}b_{j\beta}+n_{bj}\right]\right)\cr
& = &  \frac{1}{4} b\dg_{j\alpha}b_{j\alpha} b_{j\beta}b\dg_{j\beta},
\eea
where $n_{bj} = \sum_\alpha b\dg_{j\alpha} b_{j\alpha}$ is the number of bosons on a site $j$.  The last equality is due to the vanishing of antisymmetric combinations of bosons, $\tilde{\alpha}b\dg_{j\alpha}b\dg_{j-\alpha}$ on site.  Thus, for symplectic $N$, the Casimir is given by
\begin{equation}
\hat \calS_j ^{2} = \frac{1}{4}n_{bj} \left(n_{bj} +N\right),
\end{equation}
and is set by fixing the number of bosons on each site.  If we choose the convention $n_{bj} = N S$, the constraint becomes
\[
\hat \calS_j^2 = \frac{1}{4}N^2 S(S+1).
\]

The Casimir for $SU(N)$ can be obtained similarly, using the $SU(N)$ completeness relation(\ref{SUNcomplete}) instead of (\ref{complete}):
\begin{equation}
\hat \calT_j^2 =  \frac{1}{2}\left( n_b(n_b+N)- n_b - \frac{1}{N}n_b^2\right),
\end{equation}
where we have dropped the $j$ index on $n_b$ for clarity.
Using the consistent convention $n_{b} = N S$, the $SU(N)$ constraint becomes 
\[
\hat \calT_j^2 =  \frac{1}{2}(N^2-N) S(S+1).
\]  
For $N = 2$, this reduces to $S(S+1)$ and the $SU(N)$ and $SP(N)$ Casimirs are identical,  as required.  For all other $N$, $\hat \calT_j^2$ will be larger.  This means that the antisymplectic spins, $\hat \calP_j^2 = \hat \calT_j^2 - \hat \calS_j^2$ can never be removed for any $N > 2$.  In the large $N$ limit, they are forced to have equal magnitudes: $\hat \calP_j^2 = \hat \calS_j^2$.

At first sight, this requirement is quite strange.  After all, there are $N^2-1$ independent $SU(N)$ generators, which we have been treating as a vector, $\calT$, why can the spin not point in $N^2-1$ directions?  The answer is that not all directions of the $SU(N)$ vector give rise to different spins.  The spin itself is given by $\frac{1}{2}b\dg_j \cdot \calT \cdot b_j$, and $b$ has $N$ components.  The constraint removes one more degree of freedom.  For a general state, $b$ is a bosonic vector, but when the spins order, 
\begin{equation}
\langle \hat \calS_j \rangle = \frac{1}{2}\langle b\dg_{j\alpha} \mathcal{S}_{\alpha \beta} b_{j\beta} \rangle = \frac{1}{2}\langle b\dg \rangle_{j\alpha} \mathcal{S}_{\alpha \beta} \langle b \rangle_{j\beta},
\end{equation}
$\langle b \rangle$ is an $N$ component complex vector, so the spin can only take on $2N-1$ different configurations.  The spins are constrained to a $2N-1$ dimensional manifold $\mathcal{M}$.

To be more mathematically precise, this manifold $\mathcal{M}$ is a ``homogeneous space''  of $SU(N)$: $SU(N)/H_x$, where $H_x$ is the ``stabilizer'' of $x$,  the subgroup which leaves an $SU(N)$ element $x$ invariant:
\begin{equation}
H_x = \{g \in SU(N) | g \cdot x = x\}.
\end{equation}
Without loss of generality we can choose $x$ to be the spin defined by $b^T = (1, 0, \dots 0)$.  Rotating $b$ by any matrix which affects only the lowest $N-2$ entries clearly leaves $x$ invariant, as does rotating the phase of the upper two entries, so $H_x = SU(N-2)$ x $U(1)$, and
\begin{equation}
\mathcal{M}_{SU(N)} = SU(N)/SU(N-2) \mathrm{x} U(1) \cong CP^{N-1}.
\end{equation}
The full $SU(N)$ spin lives on the manifold $CP^{N-1}$, while the symplectic spin $\frac{1}{2}b\dg_j \cdot \calS \cdot b_j$ lives on a $2N-1$ dimensional manifold, given by
\begin{equation}
\label{spnmani}
\mathcal{M}_{SP(N)} = SP(N)/SP(N-2) \mathrm{x} U(1).
\end{equation}
Since $\hat \calP$ is nonzero, $\mathcal{M}_{SP(N)}$ is not contained within $\mathcal{M}_{SP(N)}$; in fact, the two manifolds have equal dimension, although they are not isomorphic.  Rather, any point on $\mathcal{M}_{SP(N)}$ will correspond to a point on $\mathcal{M}_{SU(N)}$.  Strictly speaking, this manifold is the order parameter manifold for a long range ordered state, however, it paints a useful picture of the relationship between $SU(N)$ and $SP(N)$ spins.  Furthermore, the order parameter manifold will be essential in describing the ordered state, where, for a spiral state which completely breaks the symmetry, the number of Goldstone modes will be $2N-1$.

\subsection{Ground States}

A generic Heisenberg Hamiltonian with symplectic invariance contains both antisymplectic and symplectic interaction terms, 
\begin{eqnarray}
\label{gensymp}
H &  = & \sum_{ij} J_{ij} \hat \calS_i \cdot \hat \calS_j + K_{ij} \hat \calP_i \cdot \hat \calP_j \cr
& = & \sum_{ij}\left(K_{ij}-J_{ij}\right)B\dg_{ij}B_{ij} +\left(K_{ij}+J_{ij}\right)A\dg_{ij}A_{ij} 
\end{eqnarray}
in a ratio $K/J$, which is $\pm 1$ for $SU(N)$ and $SP(N)$, respectively, and zero for symplectic-$N$.
In general, the physical, symplectic spins and and the antisymplectic spins may have different interaction strengths and signs.  

In the $S \rarrow \infty$ classical limit, the system is long range ordered, and all the bosons are condensed.  The ordered state is described by the angle between neighboring spins, $\phi_{ij} \equiv \phi_i - \phi_j$ which is 0 for a ferromagnet and $\pi$ for an antiferromagnet.  If we fix $\langle b \rangle_i = \sqrt{NS}(1,0, \ldots)^T$, we can rotate the top two coordinates of $\langle b \rangle_j$ by
\begin{equation}
R(\phi_{ij}) = \zmatrix{\cos \frac{\phi_{ij}}{2}}{\sin \frac{\phi_{ij}}{2}}{-\sin \frac{\phi_{ij}}{2}}{\cos \frac{\phi_{ij}}{2}},
\end{equation}
which makes $\langle b \rangle_j = \sqrt{NS}\left( \sin \frac{\phi_{ij}}{2}, \cos \frac{\phi_{ij}}{2}, 0,\ldots\right)$, and the two bond expectation values will be
\bea
B_{ij} & = & \langle b \rangle_j^T \langle b \rangle_i  = NS \sin \frac{\phi_{ij}}{2}\cr
A_{ij} & = &  \langle b \rangle_i^T \langle b \rangle_j = NS \cos \frac{\phi_{ij}}{2}
\eea
Thus the ground state energy for (\ref{gensymp}) is
\begin{equation}
E = \sum_{ij} (K - J)_{ij} \sin^2 \frac{\phi_{ij}}{2} + (K+J)_{ij} \cos^2 \frac{\phi_{ij}}{2}
\end{equation}

The three special cases of interest are
\bea
\label{classicalenergy}
E_{symp-N} = \sum_{ij} N S^2 J_{ij} \cos \phi_{ij}\cr
E_{SP(N)} = \sum_{ij}-  N S^2 J_{ij}\sin^2 \frac{\phi_{ij}}{2} \cr
E_{SU(N)} = \sum_{ij}  N S^2 J_{ij}\cos^2 \frac{\phi_{ij}}{2}.
\eea
We see that for antiferromagnetic bonds in $SU(N)$, the ground state energy is zero, identical to that of the paramagnet with $\langle S \rangle = 0$, and similarly for the ferromagnetic bonds in $SP(N)$.  Only symplectic-$N$ has well defined ground states for both signs of $J$.

\fg{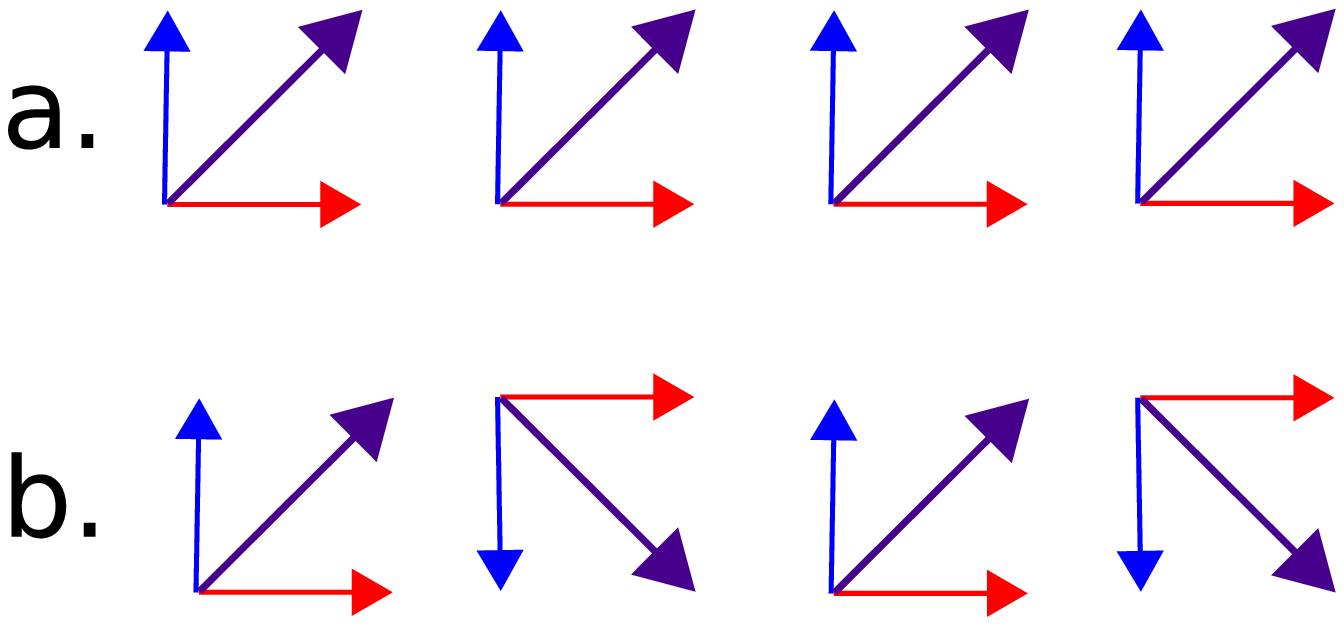}{killstates}{(Color online)A toy picture of $SU(N)$ spins, where the symplectic, time reversing components are represented in blue along the $\hat y$ axis, and the antisymplectic, non time reversing components are in red, along the $\hat x$ axis.  The purple spin shows the full $SU(N)$ spin obtained by adding its symplectic and antisymplectic components.   (a) depicts a ferromagnetic state.  (b) depicts an antiferromagnetic state, where we obtain the antiferromagnet by time reversing every other spin.  While the symplectic components are antiparallel, the antisymplectic components are still aligned, causing the total spins to be orthogonal at neighboring sites.}

If we turn to finite $S$, we can construct ferromagnetic and antiferromagnetic states explicitly out of the $SU(N)$ spins, see Fig \ref{killstates}.  The antiferromagnetic state is defined by dividing the spins into two ferromagnetic sublattices, where sublattice B spins are the time reverse of A.  This state satisfies the lattice translation plus time reversal symmetry of the $SU(2)$ antiferromagnetic ground state.  The $\hat \calP$'s are aligned in both ground states, and in the large $N$ limit, the magnitudes of $\hat \calS$ and $\hat \calP$ are the same.
In $SU(N)$, $K = J$, so both interactions are maximally satisfied in the ferromagnet - leading in fact to overstabilization due to excess $\hat \calP$ bonds, while the antiferromagnet consists of orthogonal $SU(N)$ spins, the two terms in the Hamiltonian cancel and the ground state energy is zero, just as found classically.  When $K = -J$, as in $SP(N)$, it is the antiferromagnetic ground state that is overstabilized by ferromagnetic $\hat \calP$ bonds, and the ferromagnetic state has zero energy.
These conclusions hold not only for the full ground state, but for individual bonds; in frustrated lattices there will be both antiferromagnetic and ferromagnetic correlations, even if all $J$'s are positive, but $SP(N)$ indicates ferromagnetic correlations only by the absence of a bond.  The energy cost of ferromagnetic correlations is zero in $SP(N)$, but we know that in real lattices these frustrated bonds carry a price.  By eliminating the antisymplectic interactions, symplectic-$N$ removes the extraneous bonds between $\hat \calP$'s and restores the ability of $SU(2)$ to simultaneously treat both ferromagnetism and antiferromagnetism.

\subsection{Spin dynamics}

Even if the ground state is correct, as for the bipartite antiferromagnet in $SP(N)$, we still need to be concerned about the spin dynamics.
We chose to use the group $SP(N)$ not only because its spins all invert under time reversal, but because the group contains well defined particle-particle singlets.  The presence of antisymplectic interactions, \emph{even if they are only interacting with themselves} dynamically violates the closure of the symplectic subgroup. 

The dynamics of a symplectic spin component at a site $i$ are given by
\begin{equation}
\frac{d \hat \calS_i^a}{d t} = \frac{i}{\hbar} \sum_{kj}\left( J_{kj} \left[\hat \calS_i^a, \hat \calS_k \cdot \hat \calS_j\right] +  K_{kj}\left[\hat \calS_i^a,\hat \calP_k \cdot \hat \calP_j\right]\right).
\end{equation}
We concentrate on the effect of the second term, which is generally nonzero when $K$ is nonzero.  Inserting the Schwinger boson representation, we find 
\bea
\left[\hat \calS_i^a, \hat \calP_k \cdot \hat \calP_j\right]\!\! & = &\!\!\frac{1}{8}\!\! \left[ b_i\dg \!\cdot\! \calS^a\! \cdot\! b_i, 
\left(b_k\dg\! \cdot\! \calP^b\! \cdot\! b_k\right) \left(b_j\dg\! \cdot\! \calP^b\! \cdot\!  b_j\right)\right] \cr
\!\!\!& = &\!\! \frac{1}{8}\!\{ b_j\dg \!\cdot\! \calP^b\! \cdot\! b_j, 
\left[ b_i\dg\! \cdot\! \calS^a\! \cdot\! b_i, b_k\dg\! \cdot\! \calP^b\! \cdot\! b_k 
\right] \}
\eea
where $\{,\}$ denotes the anticommutator.  Expanding out the commutator in more detail,
\bea
\left[ b_i\dg \cdot \calS^a \cdot b_i, b_k\dg \cdot \calP^b \cdot b_k 
\right] & = & \calS^a_{\alpha \beta} \calP^b_{\lambda \eta} \left[b\dg_{i\alpha}b_{i\beta}, b\dg_{k\lambda}b_{k\eta}\right] \cr
& = & \delta_{ik} b\dg_{i\alpha}\left[\calS^a,\calP^b\right]_{\alpha \beta} b_{i\beta}\cr
& = & 2 i \delta_{ik} g^{ab}_{\quad c} \hat \calT^c_{i}.
\eea
where $g^{ab}_{\quad c}$ is the appropriate $SU(N)$ structure factor.   Since the commutator, $\left[ \calS, \calP \right]$ is odd under time reversal, $\calT^c_i$ must be an antisymplectic spin.
So the evolution of $\calS_i$ is affected by the antisymplectic spins,
\begin{equation}
\left(\frac{d \hat \calS_i}{dt}\right)_{\hat \calP \cdot \hat \calP} = - \frac{1}{\hbar} \hat \calP_i \times \sum_{j}K_{ij} \hat \calP_j
\end{equation}
where $\times$ is the cross product defined by $g^{ab}_{\quad c}$.  The full dynamics of the symplectic spins are given by
\begin{equation}
\frac{d \hat \calS_i}{dt} = -\frac{1}{\hbar}\left( \hat \calS_i \times \sum_{j}J_{ij} \hat \calS_j+\hat \calP_i \times \sum_{j}K_{ij} \hat \calP_j\right).
\end{equation}
These dynamics are identical in form to classical spin wave theory, where the spins are torqued by an effective magnetic field coming from neighboring spins.  The
symplectic and antisymplectic components of $\hat \calT_i$ are torqued by the effective magnetic fields given by $\sum_{j}J_{ij} \hat \calS_j$ and $\sum_{j}K_{ij} \hat \calP_j$, respectively.  The effective field coming from the antisymplectic components is not strictly a magnetic field, as it has even time reversal parity, but most importantly, it rotates $\hat \calP$ into $\hat \calS$, and vice versa.  Ordinary $SU(2)$ spin waves will also break spin singlets, but the excitations remain in the $SU(2)$ space, while the excitations for $K \neq 0$ will take us out of the $SP(N)$ group.  It is clear that to have a theory of interacting $SP(N)$ spins, all the antisymplectic interactions must be eliminated; all other Hamiltonians with symplectic invariance describe anisotropic $SU(N)$ spin interactions.

\fg{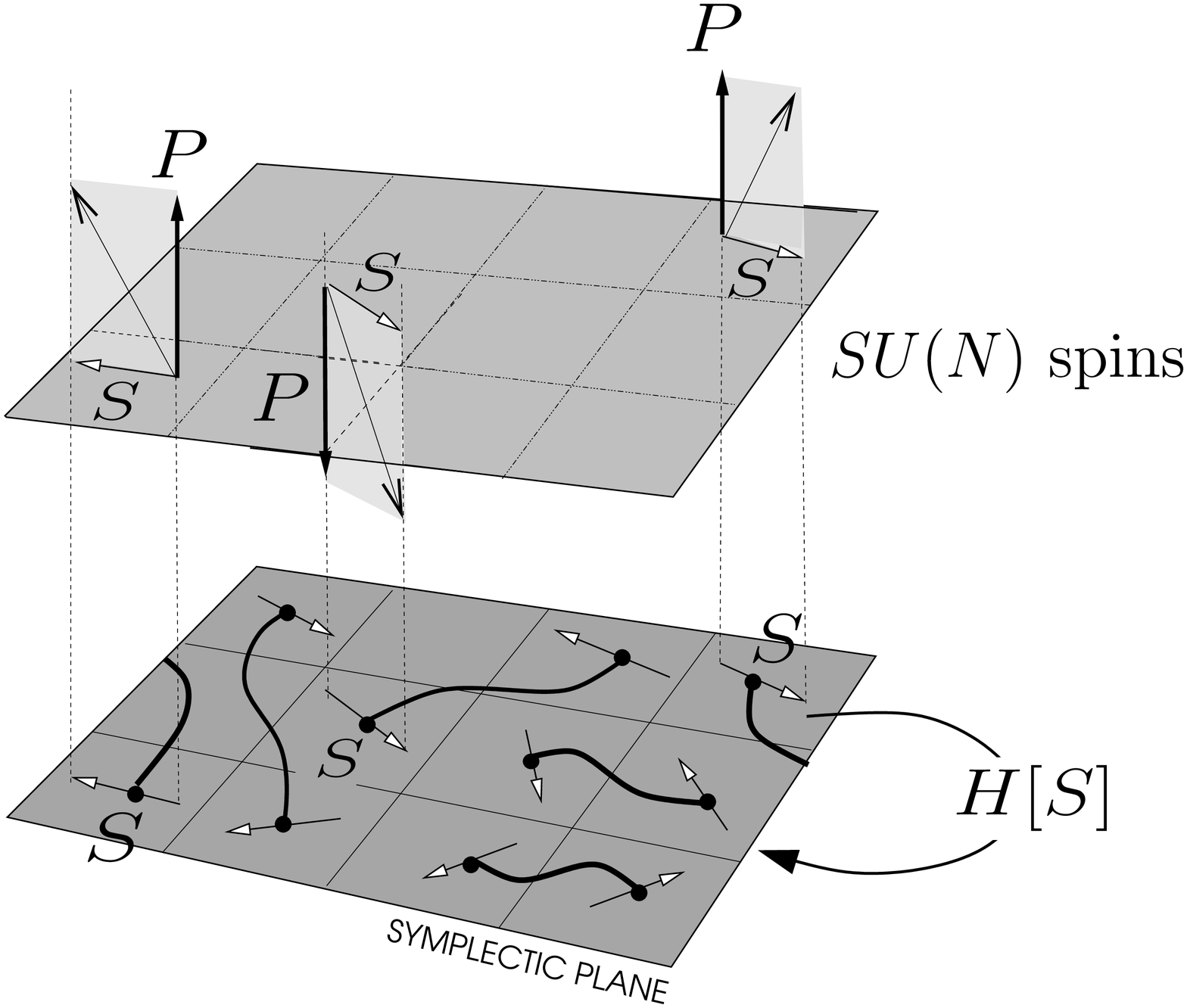}{SympPlane}{$SU(N)$ spins consist of two components: symplectic directions that reverse under time reversal, and antisymplectic directions that are invariant under time reversal, which prevent $SU(N)$ spins from forming two particle singlets.  However, if the spins are projected into the symplectic plane, these components \emph{can} form two particle singlets, which are well defined as long as the antisymplectic components are noninteracting.}

So we have seen that the inclusion of antisymplectic spin interactions have a rather serious effect on the physics of the Heisenberg model.  When these interactions are excluded, as in symplectic-$N$, the unphysical antisymplectic spins can no longer affect the physical spins.  In  a sense, they come along for the ride, since they are always there, and they are affected  by the symplectic spins, but have no effect on the physics.  Now we move on to the application of symplectic-$N$ to general lattices.

\section{Solving the symplectic-$N$ Heisenberg model}\label{GenSympH}

Now we return to the symplectic-$N$ Heisenberg model(\ref{spinham}) to discuss how to solve the Hamiltonian in the large $N$ limit, for a general lattice specified by $J_{ij}$.  As a refresher, the Hamiltonian is
\begin{equation}
\label{fullH}
H[b] = \sum_{ij} \frac{J_{ij}}{N} \hat \calS_{i}\cdot \hat \calS_{j} =
\sum_{ij}\frac{J_{ij}}{N}\left[- B\dg_{ji}B_{ji}+ A\dg_{ji}A_{ji}\right],
\end{equation}
where $B\dg_{ji} = \frac{1}{2}\tilde \sigma b\dg_{i \sigma} b\dg_{j-\sigma}$ and $A\dg_{ji}= \frac{1}{2} b\dg_{i \sigma}b_{j\sigma}$ and the sum over $\sigma$ is implied.  

The usual prescription for solving these problems is to write the partition function as a path integral,
\begin{equation}
\mathcal{Z} = \int \mathcal{D}b\,\mathrm{e}^{-N \mathcal{S}[b]} \prod_{j\tau} \delta\left( \hat \calS_j^2(\tau) - N^2 S(S+1)\right)
\end{equation}
where $N \mathcal{S}[b]$ is the action
\begin{equation}
N \mathcal{S}[b]  = \int_{0}^{\beta} d \tau \left[\sum_i \bar b_{i\sigma}(\tau) \partial_\tau b_{i \sigma}(\tau)  + H[b(\tau)]\right],
\end{equation}
and the constraint $ \prod_j \delta\left( \hat \calS_j^2(\tau) - N^2 S(S+1)\right)$ restricts the spins to the physical subspace at every site $j$ and time $\tau$.  This constraint can be rewritten using a Lagrange multiplier $\lambda_j(\tau)$,
\bea
\label{PS}
\lefteqn{\prod_{j\tau} \delta\left( \hat \calS_j^2(\tau) - N S\right) =}\cr
& & \int\!\! \mathcal{D} \lambda \exp\!\!\left[\!-\!\!\int_0^\beta\!\!\! d \tau\! \sum_j  i \lambda_j(\tau)\! \left(\bar b_{j\sigma}(\tau) b_{j\sigma}(\tau) - N S\right)\!\!\right].
\eea
From now on we drop the explicit $\tau$ dependence of $b_{i\sigma}$ and $\lambda_i$.  

In order to evaluate the path integral, $\mathcal{Z}$ must be in the form of a Gaussian integral, so the quartic terms in $H$ are decoupled using the Hubbard-Stratonovich identity,
\begin{equation}
\frac{N}{2 \pi i J}\int \mathcal{D}\Delta \, \mathrm{e}^{-N \bar \Delta \Delta/J} = 1,
\end{equation}
After inserting this identity, $\Delta$ can be shifted to $\Delta - \frac{J}{N} B$, eliminating the quartic term $\frac{J}{N} \bar B B$,
\begin{equation}
\label{tradHS}
\mathrm{e}^{\frac{J}{N} \bar B B} \propto \int \mathcal{D}\Delta \, \mathrm{e}^{-N\bar \Delta \Delta/J + \bar \Delta B +\bar B \Delta}.
\end{equation}
Now we have exchanged a theory of bosons with four particle interactions for a theory of free bosons interacting with a fluctuating field $\Delta$.  We can integrate out the bosons exactly, but we will need to use the saddle point approximation to perform the path integral over $\Delta$(Fig \ref{fluc}(a)), an approximation that becomes exact in the large $N$ limit due to the extensive dependence of the action $N \mathcal{S}$ on $N$.  First we must treat the other quartic term, $-\frac{J}{N} \bar A A$.  Naively, we would just change the sign in the exponentional in (\ref{tradHS}), which gives
\begin{equation}
\label{untradHS}
\mathrm{e}^{-\frac{J}{N} \bar A A} \propto \int \mathcal{D}h \, \mathrm{e}^{+N \bar h h/J - \bar h A -\bar A h}.
\end{equation}
However, we must be careful, as the quadratic $h$ term now has a positive sign, and the path integral over $h$ appears not to converge.  To understand this, we step back to a simpler case, where $A$ is real and we decouple it with the real field $a$.  We begin with
$\mathrm{e}^{-N a^2/J}$, and can rewrite
\bea
-N a^2/J & = & +N(i a)^2/J \rarrow +(ia + \frac{J}{N}  A)^2/J \cr
& = & -N a^2/J+ 2i A a +\frac{J}{N}  A^2,
\eea
so that the quartic term $-\frac{J}{N} A^2$ becomes $-Na^2/J +2 i A a$.  We now define the mean field value of $i a = h_0$ to be real.  In fact, let's redefine $i a = h = h_0 + i \delta a$, and the identity becomes
\begin{equation}
\mathrm{e}^{-\frac{J}{N} A^2} = \int \mathcal{D} h \mathrm{e}^{Nh^2/J + 2 A h},
\end{equation}
\fg{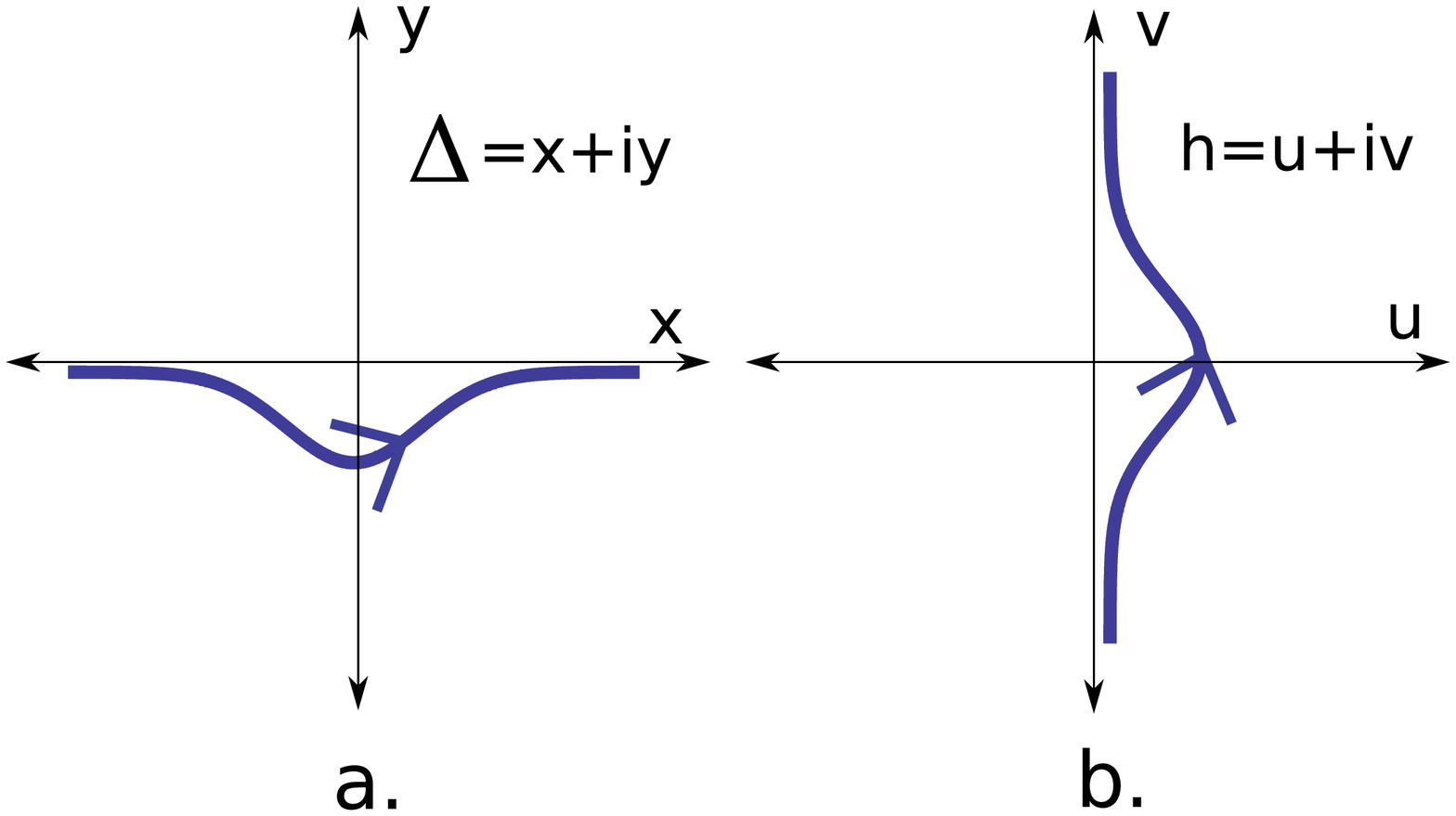}{fluc}{Integrating out the fluctuations. (a.) $\Delta$ is integrated along real axes $x$ and $y$, and its saddle point is a minima at some $\Delta_0$.  (b.)  $h$ is integrated along imaginary axes $u$ and $v$, with a maximum, real saddle point $h_0$.}
which holds as long as $h$ is integrated along the imaginary axis, with the integral maximized at a real $h_0$(see Fig \ref{fluc}(b)).  This can be generalized to a complex $a = u + i v$, where $u,v$ are imaginary instead of real.
As long as we keep in mind that $\Delta$ is integrated along the real axis and $h$ along the imaginary axis, we can proceed with the above decouplings,
\bea
\label{Hdecoupled}
H[b] = & \sum_{(ij)}& 
\left(\bar b_{i \sigma}\; \tilde\sigma b_{i-\sigma}\right)
\zmatrix{ - h_{ij}}{\Delta_{ij}}{\bar \Delta_{ij}}{-\bar h_{ij}}
\left(\!\!
\begin{array}{c}
b_{j \sigma}\cr 
\tilde \sigma \bar b_{j-\sigma}
\end{array}
\!\!\right) \cr
&+& \frac{\bar h_{ij} h_{ij} - \bar \Delta_{ij} \Delta_{ij}}{J_{ij}},
\eea
where $\sum_{(ij)}$ is performed only over bonds $(ij)$ with nonzero $J_{ij}$.  The notation can be simplified by defining the Nambu spinor, $\tilde b_j^T = \left(b_{j \sigma}, \tilde \sigma \bar b_{j-\sigma}\right)$.
We now have the partition function
\begin{equation}
\mathcal{Z} = \int \mathcal{D}\left[b,\Delta, h,\lambda\right] \mathrm{e}^{-N \mathcal{S}[b,\Delta, h,\lambda]}
\end{equation}
where the action can be compactly written
\begin{eqnarray}
\label{bAction}
N \mathcal{S}[b,\Delta, h,\lambda] \!\! & = &\!\!\!\!\!
 \sum_{i\omega_n,(ij)}\!\!\left[\,
\frac{1}{2}{\bar {\tilde b}_i}
\left(i\omega_n \tau_3+
\mathbf{G}^{-1}_{ij}\right) 
\tilde b_j \right. \cr
& + &\!\!\left.\!  \frac{N}{J_{ij}}\!\left(\bar h_{ij} h_{ij} - \bar \Delta_{ij} \Delta_{ij}\right)\! + i \lambda_i N (S + \frac{1}{2})\delta_{ij}\! \right]\cr
& & { } \quad \cr
\mathbf{G}^{-1}_{ij} & = & \zmatrix{i \lambda_i\delta_{ij} -2 h_{ij}}{2\Delta_{ij}}{2\bar \Delta_{ij}}{i \lambda_i \delta_{ij} -2\bar h_{ij}}.
\end{eqnarray}
We have performed a Fourier transform in imaginary time, and $\tilde b_i, h_{ij}, \Delta_{ij}$ and $\lambda_i$ are now functions of the Matsubara frequencies $i \omega_n$, although in practice we make the Ansatz that $h_{ij}, \Delta_{ij}$ and $\lambda_i$ are all static quantities.  The factors of $\frac{1}{2}$ come from rewriting $\lambda_i \bar b_{i\sigma} b_{i\sigma}$ in terms of the Nambu spinors, $\tilde b_i$.   

We can calculate the mean field values of $h_{ij}, \Delta_{ij}$ by approximating the path integral $\int \mathcal{D}\left[ \Delta, h, \lambda\right] \exp\left( -N S[b,\Delta, h, \lambda]\right)$ by its saddle point value, which becomes exact in the large $N$ limit.  By minimizing the action with respect to $h_{ij}$, $\Delta_{ij}$ and $\lambda_i$, we find 
\bea
\label{rsHD}
h_{ij} & = & \frac{J_{ij}}{2 N} \langle b_{i\sigma}\dg b_{j\sigma} \rangle\cr
\Delta_{ij}&  = & \frac{J_{ij}}{2 N} \langle\tilde \sigma b_{i\sigma}\dg b\dg_{j-\sigma}\rangle\cr
N S & = &  \langle b_{i\sigma}\dg b_{i\sigma}\rangle
\eea
where $\langle \cdots \rangle$ denotes the thermal expectation value.  However, it is simpler to eliminate the bosons altogether by integrating them out.  

In order to proceed further, we must make some Ansatz about $h_{ij}$, $\Delta_{ij}$ and $\lambda_i$.  In principle, $h_{ij}$ and $\Delta_{ij}$ can take different values on every bond, but for spatially uniform states, we choose an Ansatz with the unit cell of the lattice, where  $h$ and $\Delta$ are defined for each different $J_{ij}$.  If $J_{ij} = 0$ on any bond, so must $h_{ij}$ and $\Delta_{ij}$.  
We make the approximation that $i \lambda_i(\tau) = \lambda$ on every site, taking a local constraint and enforcing it only globally.  As usual, this approximation becomes exact in the large $N$ limit.

For a square lattice with only nearest neighbor couplings, this leads to three parameters, which can be further simplified to just $\lambda$ and $\Delta$, as there are no frustrating interactions.  Thus we recover the unfrustrated square lattice as previously studied in $SP(N)$\cite{readsachdev91}.  However, for frustrated lattices we cannot generally exclude either $h$ or $\Delta$.

Sometimes the uniform state will not be sufficient.  The ground state might break lattice rotational symmetries, in which case $\Delta_{i,i+\hat x}$ and $\Delta_{i,i+\hat y}$ will be different, or translational symmetry, requiring $\Delta_{i,i+\hat x} \neq \Delta_{i+\hat x, i+2\hat x}$.  When rotational symmetry is broken, $\lambda$ will remain the same on every site, but broken translation symmetry requires $\lambda_{i} \neq \lambda_{i+\hat x}$.  Since the unit cell is enlarged, there will be more than one branch of $\omega_k$, which must be summed over.  However, as long as the state may be specified by a finite number of parameters, it may be modeled within symplectic-$N$.  Problems with infinite parameter sets, e.g.- spin glasses\cite{sachdevspinglasses}, can also be treated within symplectic-$N$, but require more complicated theoretical machinery, and will not be treated here.  For the rest of this paper, we assume translational symmetry, with $i \lambda_i = \lambda$, but this treatment can be easily generalized.

The Fourier transform of the bosonic Hamiltonian, $\mathbf{G}_{ij}^{-1}$ is
\begin{equation} 
\mathbf{G}^{-1}_k = \zmatrix{\lambda - 2 h_k}{2 \Delta_k}
{\bar 2 \Delta_k}{\lambda -2 \bar h_k}.
\end{equation}
We can now perform a Bogoliubov transformation det$\left(\omega \tau_3 - \mathbf{G}^{-1}_k\right) = 0$ to obtain
\begin{equation}
\omega_k = \sqrt{(\lambda - 2 h_k)^2 - 4 \Delta_k^2},
\end{equation}
and integrate out the bosons to obtain the free energy, $F[h,\Delta,\lambda] = -\beta^{-1} \tr \log Z[b,h,\Delta,\lambda]$, where the trace is over sites $(i,j)$, and the Matsubara frequencies $i \omega_n$, in addition to the bosonic degrees of freedom.
\bea
F & = & N \beta^{-1}  \sum_k \log\left[2 \sinh \frac{\beta \omega_k}{2} \right]\cr
& + &\sum_{(ij)}\frac{N}{J_{ij}}\left(\bar \Delta_{ij} \Delta_{ij}-\bar h_{ij} h_{ij}\right) - \lambda N \mathcal{N}_s (S + \frac{1}{2}).
\eea

Let us say we have a set of $\{h_1, h_2, \ldots \}$ and $\{\Delta_1,\Delta_2, \ldots \}$, which have the Fourier transforms
\bea
h_k & =  & \sum_a h_a \gamma_{a k}\cr
\Delta_k & = & \sum_a \Delta_a \delta_{a k},
\eea
where $a$ labels a bond. The symmetry properties of $h_{ij} = h_{ji}$ and $\Delta_{ij} = - \Delta_{ji}$ force $\gamma_{a k}$ and $\delta_{a k}$ to be symmetric and antisymmetric in $k$, respectively. The free energy is now
\bea
\frac{F}{N \mathcal{N}_s} & = & \frac{\beta^{-1}}{\mathcal{N}_s} \sum_k \log\left[2 \sinh \frac{\beta \omega_k}{2} \right]\cr
& + &
 \sum_{a}\frac{z_a}{J_{a}} \left( \left|\Delta_{a}\right|^2-\left|h_{a}\right|^2 \right) - \lambda (S + \frac{1}{2})
\eea
where $z_a$ is the number of bonds of type $a$ per unit cell - for a simple square lattice this is just the coordination number $z = 4$.  The free energy is now minimized by solving the mean field equations $\partial F/\partial \lambda$, $\partial F/\partial h_a$, and $\partial F/\partial \Delta_a$:
\bea
\label{lambdaMF}
S + \frac{1}{2} & = & \frac{1}{\mathcal{N}_s} \sum_k \frac{\lambda - 2 h_k}{\omega_k}\left(n_k + \frac{1}{2}\right) \\
\label{hMF}
\frac{2 z_a h_a}{J_a} & = & - \frac{1}{\mathcal{N}_s} \sum_k \frac{\left(\lambda - 2 h_k\right) 2 \gamma_{ak}}{\omega_k}\left(n_k + \frac{1}{2}\right) \\
\label{DeltaMF}
\frac{2 z_a \Delta_a}{J_a} & = & \frac{1}{\mathcal{N}_s} \sum_k \frac{2 \Delta_k \delta_{ak}}{\omega_k}\left(n_k + \frac{1}{2}\right).
\eea
$n_k$ is the Bose function $\left(\mathrm{e}^{\beta \omega_k} -1\right)^{-1}$.

\subsection{Simple Example}

Now we examine a simple model in detail, the two dimensional bipartite square lattice.  We know the mean field value of $h$ must be zero, however, for pedagogical purposes we keep both $h$ and $\Delta$.
\begin{equation}
\omega_k = \sqrt{\left[\lambda - 2 h ( \cos k_x + \cos k_y)\right]^2 - 4 \Delta^2 (\sin k_x + \sin k_y)^2}
\end{equation}
We wish to minimize the free energy, however, we must be careful because $h$ and $\lambda$ are integrated along the imaginary axis.  In fact, the free energy should be maximized along $h$ and $\lambda$ directions and minimized along $\Delta$.  
To examine the nature of the extremum, we look at the Hessian
\begin{equation}
\bar \chi
=
\left(
\begin{array}{ccc}
\frac{\partial^2 F}{\partial \lambda^2} &  \frac{\partial^2 F}{\partial \lambda \partial h} & \frac{\partial^2 F}{\partial \lambda \partial \Delta}\\
\frac{\partial^2 F}{\partial \lambda  \partial h} &  \frac{\partial^2 F}{ \partial h^2} & \frac{\partial^2 F}{\partial h \partial \Delta}\\
\frac{\partial^2 F}{\partial \lambda  \partial \Delta} &  \frac{\partial^2 F}{ \partial h \partial \Delta} & \frac{\partial^2 F}{\partial \Delta^2}\\
\end{array}
\right)
\end{equation}
where $\Delta$ and $h$ are both zero, which is the global minimum if the temperature is well above  where $\Delta$ acquires an expectation value.  All off diagonal terms vanish at this point,
\begin{equation}
\bar \chi
=
\left(\!\!
\begin{array}{ccc}
-\frac{1}{4}\mathrm{csch}^2 \frac{\beta \lambda}{2} & 0 & 0 \\
0 & -\frac{8}{J} - \beta \mathrm{csch}^2 \frac{\beta\lambda}{2} & 0 \\
0 & 0 & \frac{8}{J} - \frac{2}{\lambda} \coth \frac{\beta \lambda}{2} \\
\end{array}
\!\!
\right).
\end{equation}
Looking at $\lambda$ and $h$ independently, $F$ is always maximized, as expected at the mean field values of $h$ and $\lambda$, while $F$ is minimized along $\hat \Delta$ for small $J$ and maximized for large $J$, indicating a second order transition to nonzero $\Delta$ at some intermediate $J$, dependent on temperature and spin.  

\subsection{Examining the Ground State}

At zero temperature, we are interested in the ground state energy,
\begin{equation}
\frac{E_0}{N \mathcal{N}_s} = \frac{1}{2\mathcal{N}_s} \sum_k \omega_k + 
 \sum_{a}\frac{z_a}{J_{a}} \left( \left|\Delta_{a}\right|^2-\left|h_{a}\right|^2 \right) - \lambda (S + \frac{1}{2}),
\end{equation}
which must again be minimized with respect to the parameters $\lambda, h_a$, and $\Delta_a$.  The order of limits is important; to obtain the correct mean field equations or $\bar \chi$, we must take the derivatives of the free energy first and then take the limit $T \rarrow 0$.  In the mean field equations(\ref{lambdaMF} - \ref{DeltaMF}), all temperature dependence is in $n_k$.  If there is no long range order, $\lim_{T \rarrow 0} n_k = 0$.  The Mermin-Wagner theorem forbids the breaking of a continuous symmetry, like $SU(2)$ or $SP(N)$ at any finite temperature in one and two dimensions\cite{merminwagner}, however at $T=0$, the Heisenberg magnet may develop long range order, which corresponds to the condensation of the Schwinger bosons\cite{hirsch}.  The bosons themselves develop an expectation value, 
\begin{equation}
\label{bCond}
b_{i\sigma} = \langle b \rangle_i + \delta b_{i\sigma},
\end{equation}
where $\overline{\langle b \rangle_i}$ and  $\langle b \rangle_i$ are no longer independent variables.  Instead $\langle b \rangle_i$ is a complex $N$ component vector, and $\overline{\langle b \rangle_i}= \langle b \rangle\dg_i$.  $n_k$ will no longer vanish for all $k$.  To see the effects of the long range order, we examine the action(\ref{bAction}) again, inserting (\ref{bCond})
\begin{eqnarray}
N \mathcal{S}[b,\Delta, h,\lambda] & = &
\!\!\int\!\! d\omega\! \sum_{(ij)}
\frac{1}{2}
\langle b \rangle\dg_i
\left(i\omega \tau_3+
\mathbf{G}^{-1}_{ij}\right)
 \langle b \rangle_j  + N S[\delta b]\cr
& = &
\frac{1}{2}\sum_{\vec Q}
\langle b \rangle\dg_{\vec Q/2} 
\mathbf{G}^{-1}_{k=\vec Q/2}
 \langle b \rangle_{\vec Q/2}
+ N S[\delta b],\quad
\end{eqnarray}
The linear terms proportional to $\delta b$ must vanish, and so have been neglected.  $\vec Q/2$ are the zeroes of the Schwinger boson spectrum.  For ferromagnetism, $\vec Q/2 = (0,0)$, while for antiferromagnetism, $\vec Q/2 = (\pi/2,\pi/2)$.  The long range order is indicated  by the ordering of the \emph{spins}, which are the combination of two Schwinger bosons, so the Goldstone modes in classical spin wave theory will be given by $\vec Q/2 \pm \vec Q/2 = \vec 0$ and $\vec Q$, which gives the traditional $(\pi, \pi)$ ordering vector for antiferromagnetism.
Now, in addition to the mean field equations, we have the condition
\begin{eqnarray}
\partial \mathcal{S} / \partial \langle b \rangle_{\vec Q/2} & = & G_{\vec Q/2}^{-1} \langle b \rangle_{\vec Q/2} = 0\cr
& = & \omega_{\vec Q/2} \langle b \rangle_{\vec Q/2} = 0.
\end{eqnarray}
So either $\langle b \rangle_{\vec Q/2} = 0$, and we proceed as before, or $\omega_{\vec Q/2} = 0$, which allows us to find the value of $\langle b \rangle_{\vec Q/2}$ in addition to the original parameters.  In fact, $n_{\vec Q/2} = \omega_{\vec Q/2} \langle b\rangle_{\vec Q/2}^2$, so we can simply define $n = n_{\vec Q/2}/\omega_{\vec Q/2}$ and the mean field equations become
\bea
\omega_{\vec Q/2}\langle b \rangle_{\vec Q/2}\!\! & = &\!\!0\cr
\label{lambdaMF0}
S + \frac{1}{2}\!\! & = & \!\!\frac{1}{\mathcal{N}_s}\! \sum_k\! \frac{\lambda - 2 h_k}{2\omega_k} 
+ \sum_{\vec Q} n(\lambda - 2 h_{\vec Q/2}) \\
\label{hMF0}
\frac{2 z_a h_a}{J_a}\!\! & = &\!\! \frac{1}{\mathcal{N}_s}\! \sum_k\! \frac{\left(2 h_k-\lambda\right)\! \gamma_{ak}}{\omega_k}\cr
& + & 2\! \sum_{\vec Q}\! n \gamma_{a \vec Q/2}(\lambda - 2 h_{\vec Q/2})\\
\label{DeltaMF0}
\frac{2 z_a \Delta_a}{J_a}\!\! & = &\!\! \frac{1}{\mathcal{N}_s}\! \sum_k\! \frac{2 \Delta_k \delta_{ak}}{2\omega_k}
+ 2 \sum_{\vec Q} n \delta_{a \vec Q/2} \Delta_{\vec Q/2}.
\eea

Now we have set up all the machinery for solving the symplectic-$N$ Heisenberg model on a general one or two dimensional lattice(three dimensional lattices cannot currently be treated by Schwinger bosons\cite{desilva}).  Next we treat a simple example which highlights the differences between symplectic-$N$ and previous large $N$ treatments, the $J_1-J_2$ model.

\section{Illustration: $J_1-J_2$ model}\label{sectionJ1J2}

The $J_1-J_2$ Heisenberg model is one of the simplest two dimensional frustrated magnets,
\begin{equation}
\label{J1J2Hamiltonian}
H = J_{1}\sum_{\bx , \mu}  \vec{S}_\bx  \cdot \vec{S}_{\bx +\mu}+ 
J_{2}\sum_{\bx ,\mu'}\quad \vec{S}_{\bx }\cdot \vec{S}_{\bx +\mu'},
\end{equation}
where $J_1$ and $J_2$ describe nearest and next nearest neighbor interactions, respectively.  We consider only antiferromagnetic $J_1$ and $J_2$.

\fg{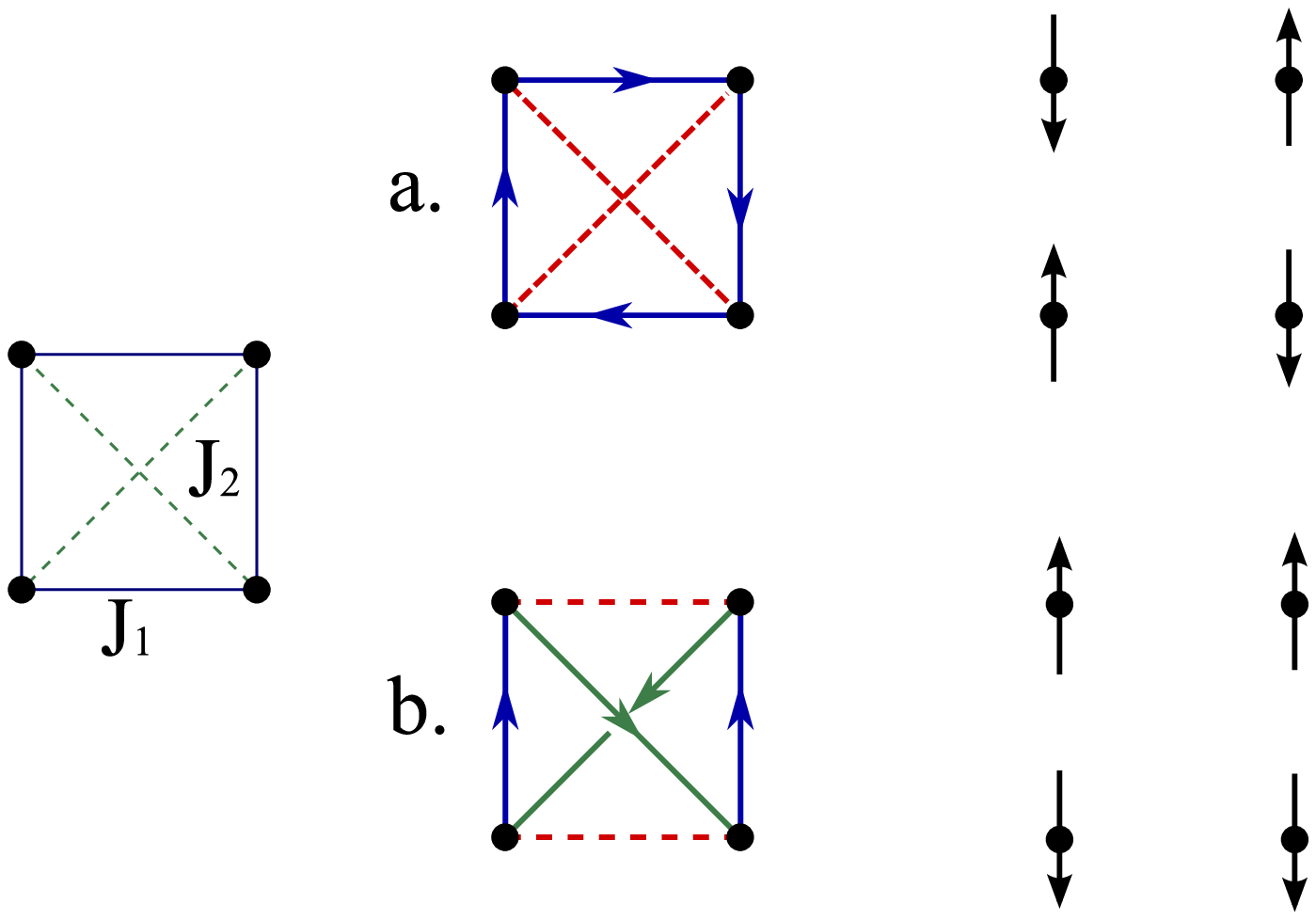}{J1J2def}{(Color online)The $J_1-J_2$ model.  (a) depicts antiferromagnetic order as described by antiferromagnetic valence bonds(blue) and ferromagnetic bonds(red, dashed), and the spin order $\vec Q = (\pi,\pi)$. (b) depicts the collinear order, $\vec Q = (0,\pi)$, where there are two different antiferromagnetic valence bonds(blue and green) and ferromagnetic bonds(red, dashed).}

For $J_1 \gg J_2$, the ground state is a N\'{e}el antiferromagnet, with $\vec Q = (\pi,\pi)$ long range order, as long as the spin $S$ is greater than a critical spin $S_c \approx .2$.  The next nearest neighbors are ferromagnetically aligned, so $J_2$ introduces frustration which begins to surpress long range order by increasing the critical spin.  

For $J_2 \gg J_1$, the classical ground state consists of two interpenetrating but decoupled N\'{e}el sublattices.  For any finite $J_1$, both quantum and thermal fluctuations couple the sublattices together through the process of ``order from disorder,''\cite{villain,shender,henley} which leads to a long range ordered state with $\vec Q = (0,\pi)$ or $(\pi,0)$.  This transition spontaneously breaks the $Z_4$ lattice symmetry down to $Z_2$, which, as a Ising symmetry  breaking, can survive to finite temperatures, despite the loss of the underlying long range magnetic order\cite{ccl}.  In real materials, this transition couples to the lattice and causes a structural transition from tetragonal to orthorhomic symmetry\cite{J1J2lattice}.

The phase boundary between the two classical ground states is at $J_1 = 2J_2$. Conventional spin wave theory predicts that the ordered moments of both states are suppressed to zero even for $S \rarrow \infty$ at this critical point, leaving a quantum spin liquid state that exists for a small, but finite range of $J_2/J_1$ for the physical spin $S=1/2$\cite{chandradoucot}.  However, at this point, the $1/S$ expansion fails\cite{xu}, and much more theoretical work has been done to see if quantum fluctuations stabilize or destabilize the spin liquid region\cite{trumper,capreview,kruger}.  The current consensus is that the spin liquid ground state is most likely stable between $.4 \lesssim J_2/J_1 \lesssim .6$ for $S=1/2$.
 
The $J_1-J_2$ model is an ideal demonstration of the importance of ferromagnetic bonds because they enforce the frustration price in both N\'{e}el and collinear phases.  This is most obvious on the N\'{e}el side, where, without ferromagnetic bonds, the state remains unchanged as $J_2$ increases, until a first order transition to the collinear state.  The ferromagnetic bonds also enable us to obtain the correct temperature dependence of the Ising transition temperature in a large $N$ theory.

\subsection{Valence bond structure}

First we need to describe the relevant states within our valence bond picture.  We assign a $\Delta$ to each antiferromagnetic bond and an $h$ to each ferromagnetic bond.  On the N\'{e}el side, we have $\Delta$ on all nearest neighbor bonds and $h_d$ on the frustrating diagonal bonds, as shown in Figure \ref{J1J2def}(a) .  This leads to the dispersion relation:
\begin{equation}
\omega_k^{n} = \sqrt{(\lambda - 4 h_d c_x c_y)^2 - 4 \Delta^2(s_x+s_y)^2)}.
\end{equation}
In the collinear state, we must allow the breaking of lattice symmetry and consider both $h_x, h_y$ and $\Delta_x, \Delta_y$ on the nearest neighbor bonds, and $h_d$, $\Delta_d$ on diagonal bonds.  In fact, there are two distinct diagonal bonds corresponding to what would be the two decoupled sublattices(see Fig \ref{J1J2def}(b)).  Their magnitude must be the same, but the phase between them leads to a $U(1)$ gauge symmetry.  If we fix the phase to be $\pi$ it is most natural to break the lattice symmetry explicitly\cite{note_gauge} and choose only $h_x$ and $\Delta_y$ to be nonzero of the nearest neighbor bonds, and $h_d = 0$, which gives the dispersion  
\begin{equation}
\omega_k^{c} = \sqrt{(\lambda - 2 h_x c_x)^2 -  (4 \Delta_dc_x s_y+2 \Delta_y s_y)^2)}.
\end{equation}

\subsection{T=0 phase diagram}

To examine the  frustrating effects of the ferromagnetic bonds, we focus on the border between long and short range orders at $T = 0$, as a function of spin, $S \equiv n_b/N$ and frustration, $J_2/J_1$ .  The more stable the phase, the larger the region of long range order.    Long range order is lost as the spin decreases below a critical spin $S_c$, which is approximately $1/5$ for the unfrustrated N\'{e}el lattice, e.g. both $J_2/J_1 =0$ and $J_2/J_1 = \infty$.  To compare our results to the original $SP(N)$\cite{readsachdev91long}, we calculate the phase boundaries both with $h$ free and with $h$ set to zero.  

Since we are interested in $S_c$, the onset of long range order, we know that both $n = 0$ and $\omega_{\vec Q/2} = 0$.  The Schwinger boson gap is at $\vec Q/2 = (\pi/2,\pi/2)$ for the N\'{e}el phase, and $\vec Q/2 = (0,\pi/2)$ for the collinear state.  The gap equation, $\omega_{\vec Q/2} = 0$ and the mean field equations for $h$(\ref{hMF0}) and $\Delta$(\ref{DeltaMF0}) can be used to solve for the mean field parameters, and  then equation (\ref{lambdaMF0}) defines $S_c$,
\begin{equation}
S_c + \frac{1}{2} = \int_k \frac{\lambda - 2 h_k}{2 \omega_k}.
\end{equation}

Results from these calculations for both symplectic-$N$ and $SP(N)$ are shown in Fig \ref{ZeroT}.  For comparison, we have also drawn the phase boundaries given by conventional spin wave theory\cite{chandradoucot}. 
\fg{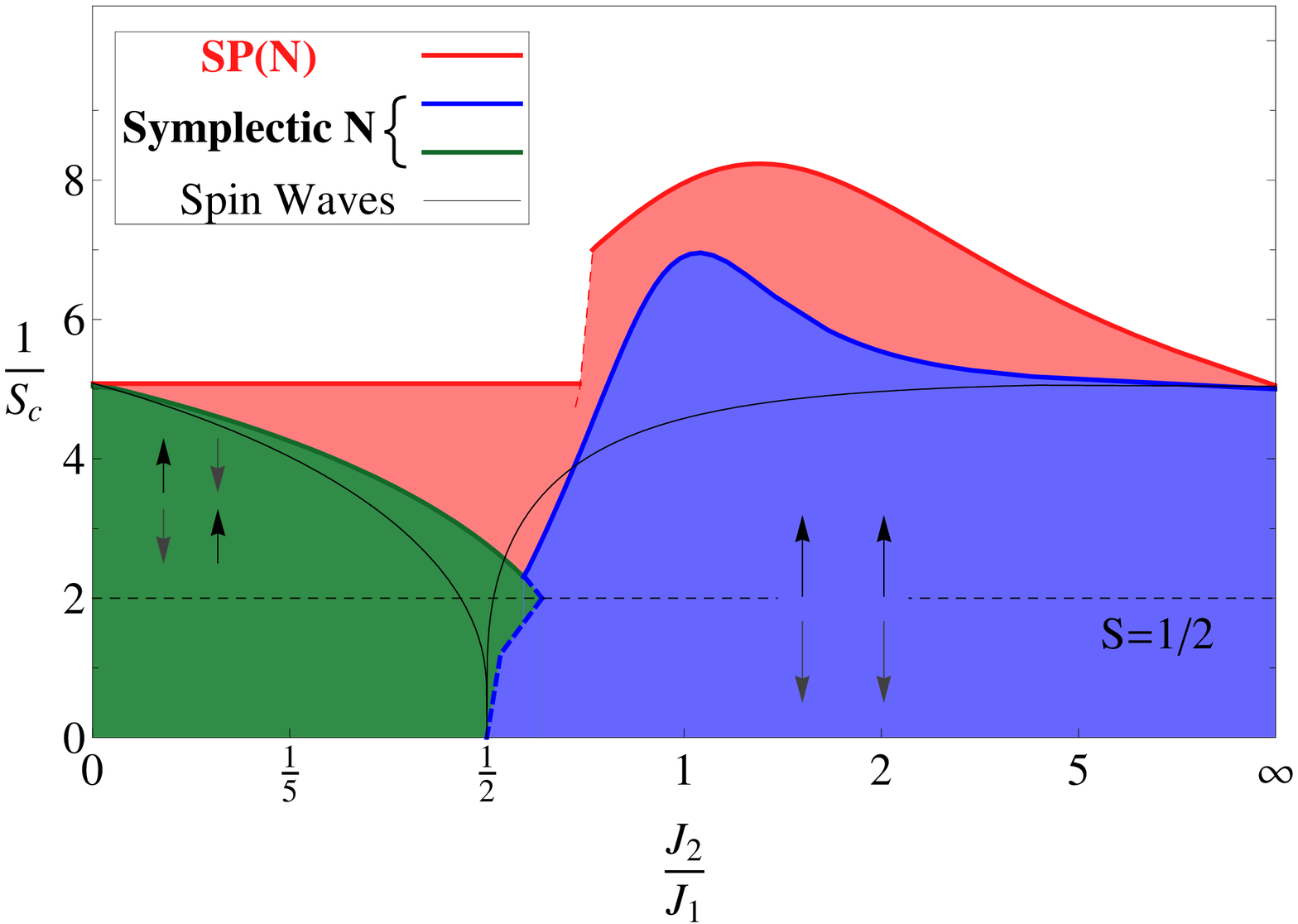}{ZeroT}{(Color online)We
compare the critical spin  $S_c = \left(\frac{n_{b}}{N}\right)_c$,
below which there is no long range order in the ground state, calculated within
$SP(N)$ (bold red line), symplectic-$N$ (blue and green lines), and spin wave
theory\cite{chandradoucot} (thin black line).  For small
$J_2/J_1$, the spins configurations are staggered, 
while for large $J_2/J_1$, the
ground state breaks lattice symmetry to develop 
collinear order as shown in the figure.  $SP(N)$ (bold red line)
tends to overstabilize the long range ordered phases, most
dramatically on the one sublattice side, where the critical spin is
independent of the strength $J_{2}$ of the 
frustrating diagonal bonds\cite{oleg}.  Symplectic-$N$
restores the frustration-induced fluctuations by treating both
ferromagnetic and antiferromagnetic bonds, on equal footing, which
corrects this overstabilization.  The physical spin, $S = 1/2$ is indicated by a horizontal dashed line.}
First let us discuss the results far from the critical value of frustration $J_1 \approx 2 J_2$.  The results are most dramatic for the N\'{e}el state, where $SP(N)$ is oblivious to the frustrating effects of the diagonal bonds, drastically overestimating the critical spin.   For the collinear state, $SP(N)$ neglects the frustrating $h_x$, again overestimating the stability of the long range ordered state.  On the other hand, symplectic-$N$ tracks conventional spin wave theory for small amounts of frustration, but they differ in a wide range around the critical $J_2/J_1$, where conventional spin wave theory is known to fail, and a spin liquid ground state is predicted for $S=1/2$.  For symplectic-$N$, we calculated the location of the first order transition between N\'{e}el and collinear long range orders by comparing the ground state energies of both states(see Appendix \ref{AppFirst}).  Symplectic-$N$ indicates a weakly first order transition for $S = 1/2$, with no intervening quantum spin liquid, however, $1/N$ corrections, calculated as Gaussian fluctuations from Ceccatto et al.'s mean field theory lead to a small region of spin liquid for $.53 \leq J_2/J_1 \leq .64$\cite{trumper}.

\subsection{Finite Temperatures: the Ising transition}

Now we turn our focus to the $J_2 \gg J_1$ side of the phase transition and examine the finite temperature Ising transition between decoupled sublattices and the collinear phase.  This phase transition has several possible experimental realizations, most prominently and recently in the iron arsenides\cite{si, haule, yildirim, muller, delacruz}.  

At high temperatures and large $J_2/J_1$, the first bonds to develop are the diagonal bonds, $\Delta_d$.  From the mean field equations(\ref{lambdaMF}) and (\ref{DeltaMF}), 
\begin{eqnarray}
\label{decoupled}
(S+\frac{1}{2})  & = &  \int_k (n_k + \frac{1}{2}) \frac{\lambda}{\omega_k} \\
\frac{1}{J_2} & = & \int_k(n_k + \frac{1}{2}) \frac{2 (2 c_x s_y)^2}{\omega_k}
\end{eqnarray}
we can solve for $\lambda$ and $\Delta_d$ as functions of temperature and spin.  They are both independent of $J_1$.  $\Delta_d$ turns on at a temperature 
\begin{equation}
\label{Td}
T_d = \frac{J_2 (S+1/2)}{2 \log (1+1/S)}
\end{equation}
Now that we have a full description of the decoupled phase, we can look for the next bond fields to turn on as we lower the temperature.  For simplicity, we assume that the spin is large enough that the ground state is the long range ordered collinear state, so we know that $h_x$ and $\Delta_y$ must turn on at some point.  However, we can look for all possible bonds at once by examining the unstable eigenvalues of the Hessian of the free energy,
\begin{equation}
\bar \chi = \left( \! \!
\begin{array}{ccc}
\frac{\partial^2 F}{\partial \lambda^2} &  \frac{\partial^2 F}{\partial \lambda \partial h_a} & \frac{\partial^2 F}{\partial \lambda \partial \Delta_a}\\
\frac{\partial^2 F}{\partial \lambda  \partial h_a} &  \frac{\partial^2 F}{ \partial h_a^2} & \frac{\partial^2 F}{\partial h_a \partial \Delta_b}\\
\frac{\partial^2 F}{\partial \lambda  \partial \Delta_a} &  \frac{\partial^2 F}{ \partial h_a \partial \Delta_b} & \frac{\partial^2 F}{\partial \Delta_a^2}\\
\end{array}
\! \! \right)
\end{equation}
where this is a schematic of the seven by seven Hessian with respect to $\lambda$, $h_x$, $h_y$, $h_d$, $\Delta_x$, $\Delta_y$, $\Delta_d$.   When $\det \bar \chi$ changes sign, the decoupled solution is changing from a free energy minimum to a maximum, indicating the presence of a second order phase transition.  By examining the unstable eigenvectors, we know which bond fields are turning on, without having to solve the seven mean field equations.

All of the matrix elements have similar forms, for example
\begin{equation}
\frac{\partial^2 F}{\partial h_x^2} = \!\!\int\!\! \frac{d^2k}{(2\pi)^2} (n_k + \frac{1}{2})\frac{\partial^2 \omega_k}{\partial h_x^2} -\frac{n_k(n_k+1)}{T}\!\left(\frac{\partial \omega_k}{\partial h_x}\right)^2 - \frac{1}{J_1}
\end{equation}
Since $\lambda$ and $\Delta_d$ are independent of $J_1$, we can fix $J_2 = 1$, $S = 1/2$ and easily evaluate $\bar \chi$ for all $J_1$ at a given $T$, since the integrals are all independent of $J_1$.  $\det \bar \chi = 0$ can then be solved for $J_{1c}$ and the phase transition, $T_c$ mapped out parametrically, as shown in Fig \ref{IsingFig}.   The unstable eigenvector is
\begin{equation}
\phi = \left( \begin{array}{c} -h_x \\ \Delta_y\end{array}\right),
\end{equation}
showing that the system does develop long range Ising order.
This method finds all possible second order phase transitions, however, it is blind to first order phase transitions.  As we see in the figure, there is a temperature dependent first order transition between the short range N\'{e}el and decoupled orders which cuts off the second order line(see Appendix \ref{AppFirst} for derivation). 

\vskip 10mm

\fg{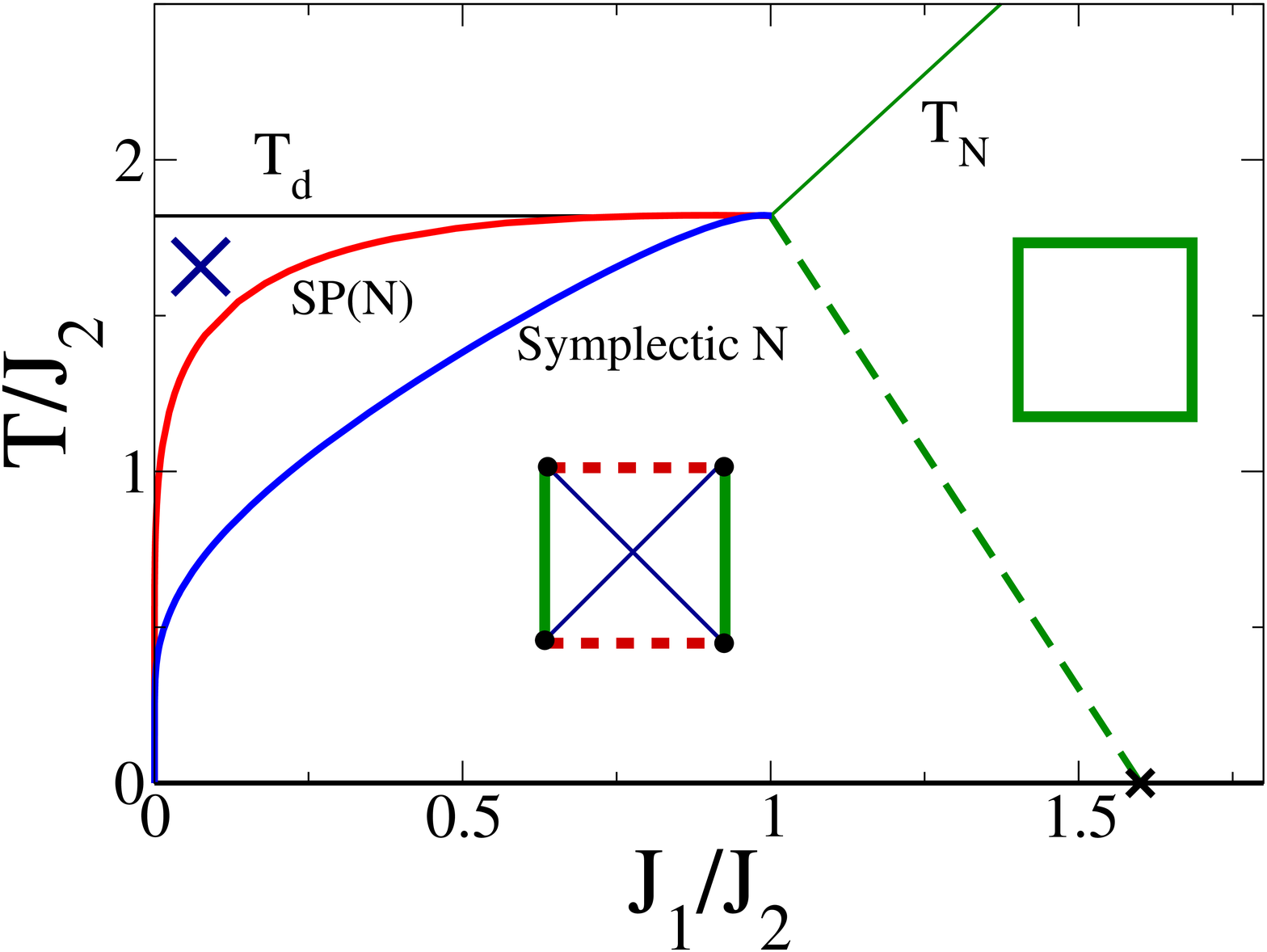}{IsingFig}{(Color online)Finite temperature phase diagram for $S = 1/2$.  $T_d$(equation \ref{Td}) and $T_N$(equation \ref{TN}) are the transitions into short range two sublattice and N\'eel antiferromagnetic order, respectively.  The Ising transition, $T_c$ is shown for both symplectic-$N$(blue) and $SP(N)$(red).  The Ising order is long range, even though the underlying antiferromagnetic order is not.  The dashed(green) line indicates a first order transition from Ising order to short range antiferromagnetic order.  Just as we saw by examining $S_c$, $SP(N)$ overstabilizes the Ising order.  Insets show the appropriate valence bond order.}

\subsubsection{Analytical form of $T_{RV\!B}$}

Now we derive the analytical form for the Ising transition temperature, in the limit of large $J_2/J_1$.  At temperatures far below the development of decoupled order $T_d$, but above the Ising transition, $T_c$, the gap in the spectrum at $(0, \pi/2)$
\begin{equation}
\Delta_{gap} = \sqrt{\lambda^2-(4 \Delta_d)^2}
\end{equation}
is much smaller than $T$, and, assuming large $S$, we can apply spin wave theory to this problem, which implies
$\lambda \approx 4 \Delta_d \approx c_{sw} = 4 J_2 S$.  $\bar \chi$ can be restricted to the two relevant parameters $h_x$ and $\Delta_y$, and we define the quantities $A_1$, $A_2$ and $B$
\begin{equation}
\bar{\chi} = \left(\!\!
\begin{array}{cc}
\frac{\partial^2 F}{\partial h_x^2} & \frac{\partial^2 F}{\partial h_x \partial \Delta_y} \\
\frac{\partial^2 F}{\partial \Delta_y \partial h_x} & \frac{\partial^2 F}{\partial \Delta_y^2}
\end{array}
\!\!
\right)  \equiv \left(
\!\!
\begin{array}{cc}
A_1 - \frac{1}{J_1} & B \\
B & A_2 + \frac{1}{J_1}
\end{array} 
\!\!\right)
\end{equation} 
In the limit $\Delta_{gap}\rightarrow 0$, we find that $A_1 = A_2 \equiv A = B$ to all divergent orders.  This is because our singlet bond fields are decoupled from the $S = 1$ spin waves becoming gapless.  To find $T_c$, we need to consider the short wavelength behavior which makes $A-B$ nonzero.
\begin{equation}
\label{chi}
\det \bar{\chi} = (A+B)(A-B) -1/J_1^2 = 0
\end{equation}
$A+B$ is of the order $T/\Delta_{gap}^2$, but the divergences cancel from $A-B$ and we can calculate this integral to zeroth order in $\Delta_{gap}$:
\begin{eqnarray}
A\!-\!B\! \! & = &\!\! \frac{1}{2}\left(\frac{\partial^2 F}{\partial h_x^2}+\frac{\partial^2 F}{\partial \Delta_y^2}\right) - \frac{\partial^2 F}{\partial h_x \partial \Delta_y} \cr
\!\!& = &\!\!2\lambda^2 \!\!\! \int\!\! \frac{d^2k}{(2\pi)^2}\frac{\cos^2 k_x\cos^4k_y}{\omega_k^2}\! \left(\!\frac{n_k(n_k+1)}{T} - 
\frac{n_k+\frac{1}{2}}{\omega_k}\!\right)\! \cr
\label{AmB}
\!\!& = &\!\! -\frac{1}{3T} \! \int\!\! \frac{d^2k}{(2\pi)^2}\frac{\cos^2 k_x \cos^4k_y}{1-\cos^2 k_x \sin^2k_y} \equiv - \frac{\pi \gamma}{T},
\end{eqnarray}
where $\gamma = .039$.
Altogether (\ref{chi}) gives us 
\begin{equation}\label{gapJ1}
\frac{8 \gamma}{\Delta_{gap}^2} = \frac{1}{J_1^2}
\end{equation}
We can expand the constraint equation(\ref{decoupled}) to find the gap
\begin{equation}
\frac{\Delta_{gap}}{c} = \exp \left(\frac{-8 \pi J_2 S^2}{T}\right)
\end{equation}
which, combined with (\ref{AmB}) leads us to the Ising transition temperature
\begin{equation}
T_{c} = \frac{4 \pi J_2 S^2}{\log \left[ \frac{2J_2S}{J_1 \sqrt{2 \gamma}}\right]}
\end{equation}
while Chandra, Coleman and Larkin found semiclassically\cite{ccl} 
\begin{equation}
T_i = \frac{4 \pi J_2 S^2}{\log \left[\frac{2J_2}{J_1 \sqrt{2\gamma_T}}\right]}
\end{equation}
with $\gamma_T = .318$.  Note that the form of the two temperatures is identical, with only numerical differences inside the logarithm, which are negligible for small spin.  This temperature dependence has been confirmed by classical Monte Carlo\cite{weber}, and quantum numerical studies have show that finite S systems also share the temperature dependence\cite{capriotti}.

The same calculation is much simpler in $SP(N)$ where $\bar \chi$ is a one dimensional matrix, 
$\partial^2 F/ \partial \Delta_y^2 \sim -T/\Delta_{gap}^2 + 1/J_1$, giving the defining condition
$T_c^{SP(N)}/\Delta_{gap}^2 = \gamma_{SP(N)}/J_1$.  Again inserting the gap(\ref{gapJ1}), we find an implicit equation for $T_c^{SP(N)}$
\begin{equation}
T_c^{SP(N)} = \frac{16 \pi J_2 S^2}{\log\left[\frac{16 J_2^2 S^2}{J_1 T_c^{SP(N)} \gamma_{SP(N)}}\right]}.
\end{equation}
The extra $T_c^{SP(N)}$ in the logarithm acts to increase the Ising temperature, as was also seen in our numerical calculation(Fig \ref{IsingFig}).

This phase transition has several possible experimental realizations.  First, there is a direct realization of the two dimensional $J_1-J_2$ lattice in Li$_2$VOSiO$_4$, where a transition to long range collinear order is immediately preceded by a lattice distortion from tetragonal to orthorhombic symmetry\cite{melzi}.  

In the iron arsenides, a $\vec Q = (0,\pi)$ spin density wave order develops either coincident with a tetragonal to orthorhombic structural transition, or slightly below\cite{delacruz,si,muller,yildirim}.  First principles calculations suggest that the system can be described by the $J_1-J_2$ model with $J_1/J_2 \approx 1/2$\cite{haule}, although whether the magnetism is itinerant or local moment is still controversial.

Finally, the spin dimer system, BaCuSi$_2$O$_6$\cite{sebastian} contains elements of $J_1-J_2$ physics despite being three dimensional.  The alternating  layers of dimers are ordered antiferromagnetically, but decoupled, like the $J_2$ sublattices, while the interlayer couplings are frustrated like $J_1$.  The compound can be thought of as a multilayer $J_1-J_2$ model, where the transition to three dimensionality is an Ising transition\cite{vojtarosch}.

\section{Discussion and Conclusions}\label{Conclusions}

We have identified the time reversal of spin as a symplectic symmetry and examined the consequences of maintaining this symmetry in the large $N$ limit.  In order to write a theory of symplectic spins, all interactions of the unphysical antisymplectic spins must be excluded, leading to a unique large $N$ limit which we call symplectic-$N$.  In this paper, we have examined the bosonic symplectic-$N$ Heisenberg model.  The practical consquences are to introduce two mean field parameters,
\bea
h_{ij} & = & \langle \frac{J_{ij}}{2N}\sum_{\sigma }b \dg_{j\sigma }b _{i\sigma }\rangle \cr
\Delta_{ij} & = & \langle \frac{J_{ij}}{2N}\sum_{\sigma }\tilde{\sigma } b \dg_{j\sigma }b \dg _{i-\sigma }\rangle
\eea
where $h_{ij}$ measures the ferromagnetic correlations along a bond $\{ij\}$ and $\Delta_{ij}$ the antiferromagetic correlations, and to identify the mean field theory introduced by Ceccatto et al\cite{ceccatto} for $SU(2)$ as the unique large $N$ limit.
Previous large $N$ methods had either ferromagnetism or antiferromagnetism, and the presence of both means that symplectic-$N$ can treat both ferromagnetic and antiferromagnetic states.  In frustrated antiferromagnets, this is especially important because the frustration manifests itself through the presence of ferromagnetic correlations on antiferromagnetic bonds; in these cases we call $h$ the frustration field.  Correctly accounting for the price of these frustrated bonds is essential in systems with many competing states close in energy.  

\fg{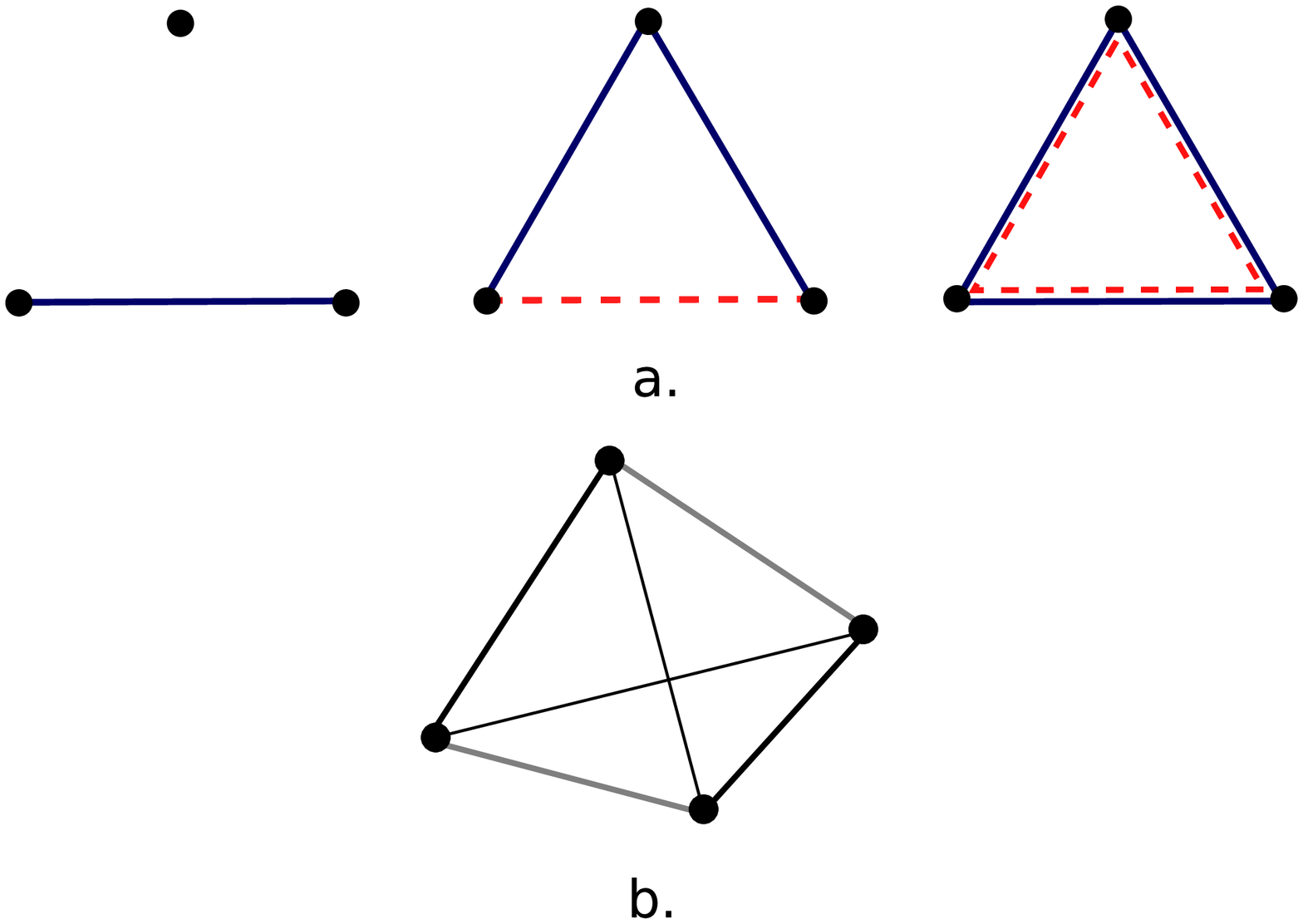}{plaquettes}{(Color online)a. The antiferromagnetic triangular plaquette has three possible bond orderings: (left) just a single $\Delta$ connecting two of the spins and leaving the third completely free; (middle) $\Delta$'s(blue) and $h$'s(red, dashed) segregated; and (right) the uniform state, which is the ground state of symplectic-$N$.b. The tetrahedral plaquette.  When all sites are assumed to be equivalent, there are three different types of bonds - as shown in black, gray and thin black lines. }

Frustrated bonds will occur whenever there are triangles containing two or more antiferromagnetic bonds(see Fig. \ref{plaquettes}(top)).  In this paper, we studied collinear magnets, where the bonds on the triangle are either exclusively ferromagnetic or antiferromagnetic, e.g. - $h$ and $\Delta$ do not coexist.  In other lattices, like the triangular lattice, noncollinear states are expected.  Certainly the classical symplectic-$N$ limit will contain coexisting bonds, as we know 
$h_{ij} = S J_{ij} \cos \frac{\phi_{ij}}{2} $, and $\Delta_{ij} = S J_{ij} \sin \frac{\phi_{ij}}{2}$, where $\phi_{ij} \neq 0$ or  $\pi$ for noncollinear ground states.  Whether this coexistence persists in the quantum limit is still an open question.  In a first attempt, we have examined the triangular plaquette and found the ground state to be the uniform, coextant state.  However, the lattice case will likely be different; the plaquette version of the $J_1-J_2$ model also has a uniform ground state, not the broken symmetry state found in the lattice.  In the tetrahedral plaquette, as in $SP(N)$, there is a continuously degenerate ground state manifold\cite{oleg}.  In $SP(N)$, the degeneracy is lifted in the lattice, however, the ground state found in the $SP(N)$ semiclassical limit is inconsistent with linear spin wave theory \cite{henleypyrochlore}.  Given the many competing states, it is an interesting open question whether the frustrating fields will bring the lattice ground state into agreement with spin wave theory.   More generally, we would like to know if including the price of frustration substantially changes the ground states or response for other highly frustrated lattices.

Now we turn to corrections beyond mean field theory, the $1/N$ corrections.  These will not affect the phase boundaries, but can change the nature of the short range phases.  Sachdev and Read have shown that the $1/N$ corrections for $SU(N)$ spins manifest as a gauge field coupling to the Schwinger bosons\cite{readsachdev89}.  In $SU(N)$, this is a $U(1)$ gauge field, which in two dimensions contains instantons that generate nontrivial Berry phases which enforce the discrete nature of valence bonds.  For each spin $S$, each site participates in exactly $2S$ valence bonds.  The ground state alternates periodically between spin-Peierls and valence bond solid phases as $2S (\mathrm{mod}\, z)$, where $z$ is the coordination number of the lattice.  Sachdev and Read later showed that this treatment can be extended to collinear states in $SP(N)$, while noncollinear states do not generally have instantons\cite{readsachdev91long}.  To examine the effects of $h$, we consider the $U(1)$ gauge symmetry in symplectic-$N$.

In the large $N$ limit of the $J_{1}-J_{2}$ model, the valence bond
fields $\Delta $ develop between ``even'' and ``odd'' sites.  This
breaks the local $U (1)$ symmetry associated with boson conservation
at each site down to a global compact $U (1)$ symmetry, under
which $b_{i}\rightarrow e^{i \theta }b_{i}$ on the even sublattice and
$b_{i}\rightarrow e^{-i \theta }b_{i}$ on the odd sublattice
(corresponding to the conservation of $\sum_{i\in \hbox{even}}n_{i}
-\sum_{i\in \hbox{odd}}n_{i}$).
The instanton tunneling configurations considered by Read and Sachdev are
space-time monopoles in the electric field associated with this $U(1)$
field. 
In fact, the frustration 
fields $h$ link sites on the same sublattice, so that $h$ is invariant
under the global $U (1)$ symmetry, so it does not pick up any phase factor 
when the instanton forms,  and it does not modify the 
the phase factors associated with instanton formation. In this way, the
frustration fields do not affect the formation of valence bond solids
in collinear states.  The effect of
the frustration fields on noncollinear states is however, still an open
question.

Another way to move beyond the large $N$ limit is to examine the variational wavefunctions which are the ground state of the large $N$ limit.  The wavefunction of a pure valence bond state has a Jastrow form\cite{liang,liquids},
\begin{equation}
\left| \Psi \right\rangle = P_S \exp\left(-\sum_{ij}b_{ij} B\dg_{ij}\right)\left| 0 \right\rangle,
\end{equation}
where $P_S$ projects out the unphysical subspace where $n_b \neq NS$, as given in equation (\ref{PS}).
When we include the effects of the frustrating fields,
\begin{equation}
\left| \Psi \right\rangle = P_S\exp\left(-\sum_{ij}a_{ij} A\dg_{ij}\right)\exp\left(-\sum_{ij}b_{ij}B\dg_{ij}\right)\left| 0 \right\rangle.
\end{equation}
The $\exp\left(-\sum_{ij}a_{ij}A\dg_{ij}\right)$ creates effective valence bonds of all lengths across the system, causing the spins to fluctuate coherently - in the case of the Ising transition, these coherent fluctuations break lattice symmetry without long range magnetic order. 

This paper has addressed the bosonic representation of interacting symplectic spins, but the principles of symplectic closure can equally well be applied to fermionic models, either in Heisenberg physics, where Ran and Wen have used an identical decoupling\cite{wen}, or Kondo physics, as we have done in collaboration with Maxim Dzero\cite{nphysus}.  In the fermionic spin representation, requiring spins that reverse under time reversal also insures that the spins are neutral under particle-hole transformations, which leads to a local $SU(2)$ gauge symmetry.  In bosonic models, this gauge symmetry reduces to the $U(1)$ symmetry discussed earlier because $\tilde \sigma b\dg_{\sigma}b\dg_{-\sigma} = 0$ on site due to symmetrization.  In parallel with our current treatment of both ferromagnetism and antiferromagnetism in the Heisenberg model, we are able to treat both the Kondo effect and superconductivity within the two channel Kondo model\cite{nphysus}. 

The next step is to introduce charge fluctuations while maintaining the symplectic spin closure.  One possibility is to introduce the symplectic-$N$ Hubbard operators, which can used to construct the $t-J$ and Anderson models.  These ensure that a hole hopping onto a site and off again will generate a symplectic spin flip.  In turn, the symplectic closure guarantees that the local $SU(2)$ gauge symmetry survives to all orders in $N$, even at finite doping, justifying the $SU(2)$ slave boson theory of Wen and Lee in a unique large $N$ limit\cite{SU2slave}.  The application of this approach as a large $N$ framework for the RVB theory of superconductivity\cite{baskaran, kotliar} is a matter of great interest for future research.

\section*{Acknowledgements}
The authors would like
to thank N. Andrei, K. Basu, M. Dzero, E. Miranda, R. Moessner, N. Read, S. Sachdev, S. Thomas and G. Zarand, for discussions
related to this work.  This research was supported by the National Science Foundation grant
DMR-0605935.  

\appendix
\section{$J_1-J_2$ first order transitions}\label{AppFirst}

In principle, calculating first order transitions is simple - one calculates the parameters $\lambda$, $h_a$ and $\Delta_a$ for each of the phases from the mean field equations, plugs them into the free energy, or ground state energy at zero temperature and compares the energies.  In practice, it is difficult to solve the mean field
equations in complicated phases.  Second order transitions are much easier because something is going to zero.
For the zero temperature phase diagram of the $J_1-J_2$ model, Fig. \ref{IsingFig}, we know that the transition between N\'eel and collinear long range order is first order because the second order lines(between short and long range order of the same type) indicate that the phases overlap for $S \gtrsim .4$.  We have calculated the location of the first order line by comparing the energy of the long range ordered states, shown in Fig. \ref{firstorder}.

\fg{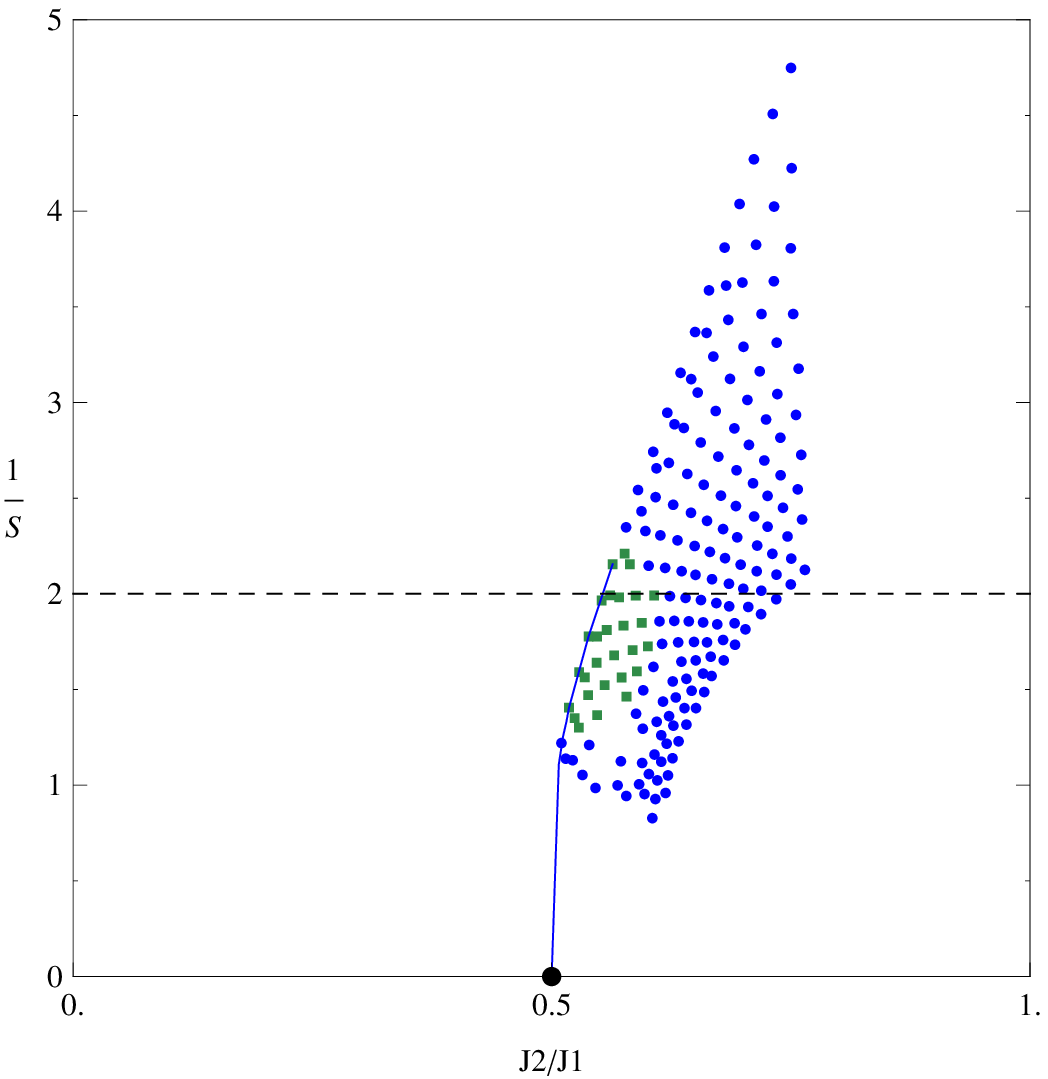}{firstorder}{(Color online)Thin lines indicate the second order transitions from short range to long range N\'eel and collinear orders.  For points within the region of collinear long range order, the ground state energies of the two possible orders were compared.  Where the collinear order is lower, a circular, blue dot is placed on the phase diagram; when N\'eel order is lower, the dot is green and square.  The black dot indicates the classical, second order phase transition, which was calculated analytically.  The physical spin, $S=1/2$ is indicated by the dashed line.}

As the spins become more classical, the mean field parameters become more difficult to calculate,  but the $S \rarrow \infty$ point can be calculated analytically using the energies from the previous section(\ref{classicalenergy}).  For the N\'eel state, $\phi_{ij} = \pi$ for nearest neighbor bonds, and 0 for diagonal bonds, while for the collinear state $\phi_{ij} = \pi$ on $\hat y$ and diagonal bonds and 0 on $\hat x$ bonds.
\bea
E_{N} & = & -4 J_1 + 4 J_2\cr
E_{c} & = & -4 J_2
\eea
Thus, the classical transition is second order at $J_1 = 2 J_2$, just as found for classical $SU(2)$ spins.  The same calculation can be repeated with $E_{SP(N)}$, with the same result.

At finite temperatures, there is a first order transition between the two short range orders.  We already have one end of the first order line - the zero temperature point, and we can calculate the other end, which is a second order point where both antiferromagnetic and decoupled short range orders give way to a completely disordered high temperature phase.  We already know the decoupled temperature as a function of $J_2$(\ref{Td}), and the antiferromagnetic temperature can be similarly found from the mean field equations for $\lambda$ and $\Delta$ in the limit of $\Delta \rarrow 0$,
\begin{equation}
\label{TN}
T_{N} = \frac{J_1 (S +1/2)}{2 \log\left(1+1/S\right)}.
\end{equation}
The two temperatures have identical form - the only difference being that where $T_N$ has $J_1$, $T_d$ has $J_2$.  Thus the first order line ends in a second order point at $J_1 = J_2$, as show in Fig \ref{IsingFig}.  The intermediate line has been extrapolated, but not calculated.

%
%
%
%


\begin{thebibliography}{99}

\bibitem{sliquid}G. Misguich,	arXiv:0809.2257v1(2008) and references therein.

\bibitem{pyrochlore} G. Misguich and C. Lhuillier, ``Frustrated spin systems'', H. T. Diep editor, World-Scientific (2005) and references therein.

\bibitem{sachdevkagome} S. Sachdev, Phys. Rev. B {\bf 45}, 12377(1992).

\bibitem{kagomeexpt}S.-H. Lee, H. Kikuchi, Y. Qiu, B. Lake, Q. Huang, K. Habicht and  K. Kiefer, Nature Materials {\bf 6}, 853 - 857 (2007).

\bibitem{oleg}O. Tchenyshyov, R. Moessner and S. L. Sondhi, Europhysics Letters {\bf 73}, 278 (2006).

\bibitem{takagi} Y. Okamoto, M. Nohara, H. Aruga-Katori and H. Takagi, Phys. Rev. Lett. {\bf 99}, 137207 (2007).

\bibitem{hyperkagomeclassical}  J. M. Hopkinson, S. V. Isakov, H.-Y. Kee, and Y. B. Kim, Phys. Rev. Lett. {\bf 99}, 037201 (2007).

\bibitem{ybkim} M. J. Lawler, H.Y. Kee, Y. B. Kim, and A. Vishwanath, Phys. Rev. Lett. {\bf 100}, 227201 (2008) 

\bibitem{qsnematic} P. Chandra and P. Coleman, Phys. Rev. Lett. {\bf 66}, 100 (1991).

\bibitem{chiralspin} X. G. Wen, Frank Wilczek, and A. Zee, Phys. Rev. B {\bf 39}, 11413 (1989).

\bibitem{witten}E. Witten, Nucl Phys. B {\bf 160}, 57 (1979).

\bibitem{read}N. Read, and D. M. Newns, J. Phys. C {\bf 16}, L1055, (1983).

\bibitem{read2} N. Read and D.M. Newns, J. Phys. C {\bf 16}, 3273 (1983).

\bibitem{auerbach}A. Auerbach, and K. Levin, Phys. Rev. Lett. {\bf  57}, 877 (1986).

\bibitem{spherical} T.H. Berlin and M. Kac, Phys. Rev. {\bf 86}, 821(1952).

\bibitem{anderson} P.W. Anderson, Phys. Rev. {\bf 86}, 694 (1952).

\bibitem{dyson} Freeman Dyson, Phys. Rev. {\bf 102}, 1217 (1956).

\bibitem{maleev}S.V.Maleev, Zh. Eksp. Teor. Fiz. {\bf 33}, 1010 (1957).

\bibitem{NLSM}S. Chakravarty, B. I. Halperin, and D. R. Nelson, Phys. Rev. Lett. {\bf 60}, 1057(1988);  Phys. Rev. B {\bf 39}, 2344(1989).

\bibitem{dadda} A. D'Adda, M. L\"uscher, and P. Di Vecchia, Nucl. Phys. B {\bf 146}, 63(1978).

\bibitem{affleckchains}Ian Affleck, Phys. Rev. Lett. {\bf 54}, 966(1985).

\bibitem{affleckmarston}I. Affleck and J.B. Marston, Phys. Rev. B {\bf 37}, 3774(1988).

\bibitem{wen} Y. Ran and X.G. Wen, cond-mat/0609620v3 (2006)

\bibitem{U1spinliquid}M. Hermele, T. Senthil, M. P. A. Fisher, P. A. Lee, N. Nagaosa, and X.-G. Wen, Phys. Rev. B {\bf 70}, 214437 (2004).

\bibitem{schwinger}D.P. Arovas and A. Auerbach, Phys. Rev. B {\bf 38}, 316 (1988).

\bibitem{readsachdev91}N. Read and Subir Sachdev,
 Phys. Rev. Lett. {\bf  66}, 1773 (1991); 

\bibitem{nphysus}R. Flint, M. Dzero and P. Coleman, Nat. Phys. {\bf 4}, 643 (2008).

\bibitem{valencebonds} P.W. Anderson, Mater. Res. Bull. {\bf 8}, 153(1973).

\bibitem{ceccatto}H.A. Ceccatto, C.J. Gazza and A.E. Trumper, Phys. Rev. B {\bf   47}, 12329 (1993).


\bibitem{sachdevspinglasses} S. Sachdev and J. Ye, Phys. Rev. Lett. {\bf 70}, 3339(1993);A. Georges and O. Parcollet and S. Sachdev, Phys. Rev. Lett. {\bf 85}, 840(2000)

\bibitem{merminwagner} N. D. Mermin and H. Wagner, Phys. Rev. Lett. {\bf 17} 1133(1966).

\bibitem{hirsch} J.E. Hirsch and S. Tang, Phys. Rev. B {\bf 39}, 2850 (1989).

\bibitem{desilva}T.N. De Silva, M. Ma, and F.C. Zhang, Phys. Rev. B {\bf 66}, 104417 (2002).

\bibitem{villain} J. Villain, J. Phys.(Paris) {\bf 38} 26(1977);  J. Villain, J. Phys.(Paris){\bf 41} 1263(1980).

\bibitem{shender}E. Shender, Zh. Eksp. Teor. Fiz. {\bf 83}, 326(1982).

\bibitem{henley}C. L. Henley, Phys. Rev. Lett. {\bf 62}, 2056 (1989).

\bibitem{ccl}P. Chandra, P. Coleman and A. I. Larkin, Phys. Rev. Lett. {\bf 64}, 88 (1990).
    
\bibitem{J1J2lattice}C. Weber, F. Becca and F. Mila. Phys. Rev. B {\bf 72}, 024449 (2005).

\bibitem{chandradoucot} P. Chandra and B. Doucot, Phys. Rev. B{\bf  38}, 9335, 1988.

\bibitem{xu} J.H. Xu and C.S. Ting, Phys. Rev. B {\bf 42}, 6861-6864(1990).

\bibitem{trumper}A.E. Trumper, L.O. Manuel, C.J. Gazza and H.A. Ceccatto, Phys. Rev. Lett. {\bf 78}, 2216(1997).

\bibitem{capreview} L. Capriotti,  Int. J. Mod. Phys. B {\bf 15}, 1799 (2001) and references therein.

\bibitem{kruger} F. Kr\"uger and S. Scheidl, Europhys. Lett. {\bf 74}, 896(2006).

\bibitem{note_gauge}We could still treat the other case(after all it's just a gauge symmetry) by using staggered bond values instead of uniform., but this would require doubling the unit cell.

\bibitem{readsachdev91long} Subir Sachdev and N. Read, Int. J. Mod. Phys. B {\bf 5}, 219(1991).

\bibitem{si} Q. Si and E. Abrahams, Phys. Rev. Lett. {\bf 101}, 076401(2008).

\bibitem{muller}C. Xu, M. Muller, S. Sachdev, Phys. Rev. B {\bf 78}, 020501(R)(2008).

\bibitem{yildirim}T. Yildirim, Phys. Rev. Lett {\bf 101}, 057010 (2008).

\bibitem{delacruz}Clarina de la Cruz, Q. Huang, J. W. Lynn, Jiying Li, W. Ratcliff II, J. L. Zarestky, H. A. Mook, G. F. Chen, J. L. Luo, N. L. Wang and  Pengcheng Dai, Nature(London) {\bf 453}, 899-902(2008).

\bibitem{haule} K. Haule and G. Kotliar, arXiv:0805.0722v1(2008).

\bibitem{weber}C. Weber, L. Capriotti, G. Misguich, F. Becca, M. Elhajal, and F. Mila, Phys. Rev. Lett. {\bf 91}, 177202(2003).

\bibitem{capriotti}L. Capriotti, A. Fubini, T. Roscilde, and V. Tognetti, Phys. Rev. Lett. {\bf 92}, 157202(2004) .

\bibitem{melzi}R. Melzi, P. Carretta, A. Lascialfari, M. Mambrini, M. Troyer, P. Millet, and F. Mila, Phys. Rev. Lett. {\bf 85}, 1318-1321(2000).

\bibitem{sebastian}S.E. Sebastian, N. Harrison, C. D. Batista, L. Balicas, M. Jaime, P. A. Sharma, N. Kawashima  and  I. R. Fisher, Nature {\bf 441}, 617 (2006).

\bibitem{vojtarosch}O. R\"osch and M. Vojta, Phys. Rev. B {\bf 76}, 180401(R)(2007).

\bibitem{henleypyrochlore}U. Hizi, P. Sharma and C. L. Henley, Phys. Rev. Lett. {\bf 95}, 167203(2005).

\bibitem{readsachdev89} N. Read and S. Sachdev, Phys. Rev. Lett. {\bf 62}, 1694 (1989).

\bibitem{liang}S. Liang, B. Doucot, and P.W. Anderson Phys. Rev. Lett. {\bf 61}, 365(1988).

\bibitem{liquids}P. Chandra, P. Coleman, and A.I. Larkin, J. Phys.: Condens. Matter{\bf 2}, 7933(1990).

\bibitem{SU2slave} X.G. Wen and P.A. Lee, Phys. Rev. Lett. {\bf 76}, 503 (1996).

\bibitem{baskaran}P.W. Anderson, G. Baskaran, Z. Zou and T. Hsu, Phys. Rev. Lett.{\bf 58}, 2790(1987).

\bibitem{kotliar}G. Kotliar and J. Liu, Phys. Rev. B {\bf 38}, 5142(1988).

%

\end{thebibliography}
\end{document}